\documentclass[11pt]{article}
\usepackage[margin=1in]{geometry}
\usepackage{amsmath}
\usepackage{graphicx}
\usepackage{tikz-cd}
\newcommand{\circnum}[1]{%
\tikz[baseline=(char.base)]{
\node[shape=circle, fill=black, text=white, inner sep=1pt] (char) {\small #1};}}
\usepackage{multicol}
\usepackage[colorlinks=true,citecolor=green,linkcolor=green,urlcolor=green, pdfborder={0 0 0}]{hyperref}
\usepackage{subcaption}
\usepackage[table]{xcolor}
\usepackage{booktabs}
\usepackage{array}
\usepackage{float}
\usepackage{stfloats}
\usepackage{flafter}
\usepackage{microtype}
\usepackage{silence}
\usetikzlibrary{arrows.meta,positioning,shapes.geometric,fit,backgrounds,calc}

\title{Agentic AI for Particle-Based Simulation: Automating SPH Workflows for Debris Flow Modeling}

\author{
  Danrong Zhang\textsuperscript{1}, Ruijia Wang\textsuperscript{2}, Chenying Liu\textsuperscript{2}, Yumeng Zhao\textsuperscript{3}\thanks{Corresponding author. Email: yzhao52@unl.edu}
  \\
  \textsuperscript{1}Independent Researcher
  \\
  \textsuperscript{2}School of Civil and Environmental Engineering, Georgia Institute of Technology
  \\
  \textsuperscript{3}Department of Civil and Environmental Engineering, University of Nebraska--Lincoln
}
\date{\today} 

\begin{document}

\maketitle

\section*{Abstract} Physics-based simulation underpins engineering analysis but remains difficult to deploy in practice due to complex setup, parameterization, and interpretation. While Large Language Model-based agentic systems have shown promise in automating engineering computing workflows, they have primarily targeted structured, mesh-based problems. We present the first agentic AI workflow for meshless simulation in computational mechanics, demonstrated on debris flow modeling using Smoothed Particle Hydrodynamics (SPH) with the software DualSPHysics. By integrating tool orchestration, multimodal inputs (text and sketches), and human-in-the-loop interaction, the framework enables end-to-end simulation workflows for a class of problems that are inherently less structured and more challenging to automate. Results show that multimodal inputs not only enhance user experience but also reduces failure modes over text-only descriptions. Human-in-the-loop is critical for resolving ambiguities and handling SPH-specific configurations. We further introduce a cognitive-task-based evaluation of post-processing, showing strong performance in visualization and data extraction, with remaining gaps in higher-level SPH-specific physical reasoning that are amenable to improvement through domain-aware modeling. These results establish the viability of agentic AI for particle-based simulation and underscore its potential to transform the accessibility and efficiency of computational mechanics workflows.\\

\textbf{Keywords} Large Language Model; Agentic Workflow; Simulation Orchestration; Particle-Based Method; Smoothed Particle Hydrodynamics; Debris Flow.

\raggedcolumns
\begin{multicols*}{2}
\section{Introduction}
Over the past decades, physical-based numerical methods have
achieved remarkable success in mechanics, including solid,
fluid, and granular systems~\cite{hughes2003finite,ferziger2002computational,poschel2005computational}. They have been widely adopted
for engineering analysis and design optimization given their versatility in modeling complex physical behavior with high fidelity~\cite{oberkampf2010verification}. Despite these strengths, two major barriers still limit their direct use in
routine engineering decision-making. First, practitioners must
possess substantial expertise in numerical modeling: they need
to understand governing physics and equations, and also make
technically sensitive choices such as mesh resolution, time
step size, solver selection, and initial and boundary
conditions to ensure convergence and reliability. Second, they
must operate complex simulation software that is often
especially challenging when tools are open-source and provide
limited graphical user interfaces. As a result, practical
decisions frequently require collaboration among multiple
specialists, increasing turnaround time, communication
overhead, and cost. These limitations motivate the need for approaches that can reduce the reliance on specialized expertise and improve the accessibility of physics-based simulation tools.

While AI-aided approaches to mechanical simulation have gained traction, most efforts focus on surrogate modeling, where data-driven models approximate the input–output behavior of high-fidelity solvers~\cite{haghighat2021physics,ganti2020data,wandel2021teaching,choi2024graph,wang2025machine}. Despite advances in deep learning, surrogate models often suffer from limited generalizability, with predictive accuracy degrading when problem geometry, boundary conditions, or governing physics deviate from the training distribution~\cite{nourbakhsh2018generalizable,eiximeno2025deep}. Moreover, their development still relies on conventional simulations for data generation and validation. As a result, high-fidelity physical-based solvers remain indispensable, and improving the accessibility and usability of these tools remains a high priority.

The rapid advancement of large language models (LLMs) have enabled the development of agentic AI systems capable of tool using, multi-step reasoning, and autonomous workflow orchestration. Rather than replacing physical-based models, LLM-based systems integrate domain knowledge, external tools, and iterative decision-making to support complex scientific and engineering workflows~\cite{GUO2026118591,ali2024physics,alexiadis2024text,feng2025openfoamgpt2}. By interpreting high-level human intent, these systems can autonomously drive simulation pipelines, from parameter setup and job configuration to post-processing. For example, agentic frameworks have been developed as a domain-aware assistant for materials science~\cite{zhang2024honeycomb}, inverse design of engineered structures~\cite{lu2025agentic}, and automation of molecular dynamics workflows~\cite{campbell2026mdcrow}, demonstrating the potential of LLM-based AI-agents to streamline multi-stage computational pipelines. However, existing applications have largely focused on domains with relatively structured workflows and well-defined inputs. In contrast, physical-based simulations involving complex multiphase and mechanical behavior are inherently less structured. In particular, particle-based methods such as Smoothed Particle Hydrodynamics (SPH) have received limited attention in agentic AI frameworks.

SPH has emerged as a powerful method for simulating complex fluid, geomechanics and other granular flow problems~\cite{violeau2012fluid,bui2020smoothed,cleary2021application,zhao2023sph,zhao2024sph}. Unlike conventional mesh-based approaches such as the finite element or finite volume methods, SPH adopts a Lagrangian, meshless formulation in which the continuum is represented by moving particles rather than a fixed grid~\cite{violeau2012fluid}. This enables natural handling of free surfaces, multiphase interactions, large deformations, and moving interfaces, making SPH particularly suitable for geophysical flows such as debris flow.

Despite these advantages, the meshless nature of SPH leads to a less structured formulation than mesh-based methods, where geometry and boundary conditions are encoded through particle distributions rather than explicitly defined. As a result, automated setup is difficult. Besides, post-processing requires extracting physically meaningful fields from irregular particle data (e.g., free surfaces and particle–boundary interactions), further challenging agentic workflows. To the best of our knowledge, LLM-driven automation has not been explored for meshless (also referred to as particle-based) methods such as SPH, where additional challenges arise in discretizing complex geometries into particles, selecting stable numerical parameters, and post-processing irregular free surfaces and particle-boundary interactions. The closest intersection between agent-based methods and meshless solvers is the work of Zhan et al.~\cite{zhan2025sphmarl}, who integrated SPH with multi-agent reinforcement learning to optimize control policies. However, these AI-agents operate at runtime, while simulation setup and configuration remain manual. Thus, it is necessary to explore agentic workflow for SPH and other meshless methods.

Existing agentic workflow approaches typically rely on either natural-language descriptions or high-quality images to specify simulation scenarios. Whereas, in practice engineers often communicate problems through a combination of sketches and brief annotations, where key spatial relationships, dimensions, and boundary layouts are conveyed visually but may remain incomplete or ambiguous. Text-only inputs struggle to capture geometry, while image-only inputs lack precise parameterization. This underscores the necessity for agentic workflows that integrate multimodal data such as natural language and visual graphs, resolve ambiguities, and iteratively refine simulation configurations.

Despite rapid progress in agentic workflows for computational mechanics, existing studies primarily evaluate performance using outcome-based metrics such as executability, success rate, or cost~\cite{wang2025openfoamgpt_cost,ni2024mechagents,xu2025cfdagent}. Consequently, these evaluations provide limited insight into system reliability, failure modes, and generalizability~\cite{xu2025cfdagent,mudur2025feabench}. This leaves a critical gap in understanding how systems reason across the heterogeneous cognitive tasks inherent to simulation—such as problem formulation, parameter inference, and error diagnosis. This lack of "task-awareness" extends to system design: workflows are often treated as monolithic processes, failing to distinguish between reasoning-intensive and deterministic stages while offering limited support for structured human intervention. These limitations underscore the necessity for a task-level perspective in both the design and evaluation of agentic simulation systems.

To address these gaps, this study introduces a human-in-the-loop agentic workflow for high-fidelity mechanics simulation built on DualSPHysics v5.0, a GPU-accelerated meshless SPH framework~\cite{dominguez2022dualsphysics}. Unlike prior work focused on grid-based solvers, the proposed system targets particle-based simulation, where setup, stability, and boundary representation pose distinct challenges. The agentic workflow integrates multimodal input, including sketch-based geometry and natural-language specifications, and adopts a stage-structured design to improve robustness and interpretability. In addition, we develop a cognitive-task-oriented evaluation framework to systematically assess AI-agent performance beyond standard end-to-end metrics. Debris flow simulation with a non-Newtonian SPH model serves as the representative application. The principal contributions of this work are as follows:
\begin{enumerate}
\item A task-aware, stage-structured agentic workflow for particle-based simulation, developed on \textit{DualSPHysics}, which explicitly distinguishes between reasoning-intensive and deterministic components, enabling reliable automation while supporting structured human intervention. To the best of our knowledge, this represents one of the first dedicated agentic frameworks designed for meshless, particle-based mechanics simulation.

\item A cognitive-task-oriented evaluation framework that characterizes AI-agent performance across distinct reasoning functions, including problem formulation, parameter inference, workflow decomposition, and error diagnosis, moving beyond conventional end-to-end success metrics.

\item A multimodal interaction interface that integrates sketch-based geometric input with natural-language specifications to translate high-level user intent into physically consistent particle-based simulation configurations.

\item A human-in-the-loop co-pilot paradigm that enables structured expert intervention at stage boundaries, allowing iterative validation and refinement while reducing manual setup effort.
\end{enumerate}

\section{Related work}
To contextualize the contributions of our work, we review recent developments in agentic workflows for computational mechanics, with a focus on fluid and solid mechanics where such systems have been most actively explored.

\subsection{Agentic Workflow in Fluid Mechanics}\label{sec:related_work_fluid}
In computational mechanics, agentic LLM workflows have advanced most rapidly in computational fluid dynamics (CFD), due in part to the transparent, text-based case structure of open-source software such as OpenFOAM, which is especially amenable to automated generation, inspection, and correction. The studies typically use LLMs to interpret natural-language requests, assemble solver-ready inputs, coordinate execution, and iteratively refine outputs through tool feedback.

Early work such as MetaOpenFOAM~\cite{chen2024metaopenfoam} introduced multi-agent, retrieval-augmented pipelines for translating user requests into executable OpenFOAM simulations. This line of development was broadened by CFDAgent~\cite{xu2025cfdagent}, which demonstrated zero-shot, end-to-end CFD automation from multimodal inputs, including natural-language descriptions and images. MetaOpenFOAM 2.0~\cite{chen2025metaopenfoam2} subsequently extended the multi-agent paradigm through chain-of-thought decomposition and iterative verification, reporting an executability score of 6.3/7 and an 86.9\% pass rate across simulation and post-processing tasks. In parallel, OpenFOAMGPT~\cite{pandey2025openfoamgpt} showed that a retrieval-augmented single-agent architecture can also support case generation and correction effectively, while OpenFOAMGPT 2.0~\cite{feng2025openfoamgpt2} advanced this idea into an end-to-end four-agent workflow with 100\% reproducibility across more than 450 runs.

Subsequent studies have shifted attention toward robustness, benchmarking, and deployment. Foam-Agent~\cite{yue2025foamagent} expanded coverage to mesh generation, post-processing, and dependency-aware task scheduling, achieving an 83.6\% success rate on a 110-task benchmark and outperforming both MetaOpenFOAM and OpenFOAMGPT baselines. ChatCFD~\cite{fan2025chatcfd} further strengthened structured reasoning and benchmark-oriented evaluation through tutorial, perturbed, and literature-derived cases. In parallel, Dong et al.~\cite{dong2025nl2foam} fine-tuned Qwen2.5-7B on 28,716 natural-language-to-OpenFOAM pairs and reported an 82.6\% first-attempt success rate, while Wang et al.~\cite{wang2025openfoamgpt_cost} evaluated lower-cost variants of the OpenFOAMGPT workflow and found that token costs could be reduced by up to two orders of magnitude relative to OpenAI o1, although more challenging cases still benefited from expert supervision.
Overall, there is a clear progression from proof-of-concept case generation to retrieval-grounded, benchmarked, and increasingly workflow-aware CFD automation. 

\subsection{Agentic Workflow in Solid Mechanics}\label{sec:related_work_solid}
In solid mechanics, existing studies have centered on synthesizing finite element analysis (FEA) with AI-agents ~\cite{ni2024mechagents,tian2024fea,mudur2025feabench,zhang2025mooseagent}. In MechAgents~\cite{ni2024mechagents}, mechanics problems are framed as a multi-agent collaboration in which specialized AI-agents handle knowledge retrieval, code generation, execution, and validation, demonstrating that AI-agent decomposition can improve robustness on classical mechanics tasks. Subsequent work has focused on collaboration design and systematic evaluation. Tian and Zhang~\cite{tian2024fea} studied how agent-role assignment affects reliability in FEA automation and showed that clearly defined, complementary roles matter more than simply increasing the number of AI-agents. FEABench~\cite{mudur2025feabench} extended the field from individual case studies to benchmark-driven evaluation by testing whether language models and AI-agents can operate COMSOL Multiphysics through its API to solve end-to-end multiphysics problems.

Application-oriented frameworks have also begun to connect AI-agents to established FEM software. MooseAgent~\cite{zhang2025mooseagent} applies multi-agent decomposition, retrieval, and iterative verification to automate parts of the MOOSE workflow, illustrating how solver-grounded architectures can reduce setup burden while retaining deterministic numerical backend. Kim et al.~\cite{kim2025can} provide an early attempt to extend this agentic framework to geomaterials, a domain characterized by multiphase behavior and complex constitutive laws.

Overall, these studies suggest that solver-grounded automation with explicit validation is a promising path for solid mechanics, but the area still lacks the breadth of benchmarks and workflow maturity now seen in CFD.

Despite these advances in Section~\ref{sec:related_work_fluid} and~\ref{sec:related_work_solid}, key gaps remain in task-level reasoning and particle-based simulation. We next introduce the methodology of the proposed framework.

\section{Methods}
\subsection{Numerical Tool: DualSPHysics v5.0}
DualSPHysics v5.0 is an open-source computational mechanics software based on SPH~\cite{crespo2015dualsphysics}. The software supports highly parallel simulations with millions of particles on both CPUs and GPUs (via CUDA), making it suitable for computationally intensive applications. A detailed introduction to SPH theory and the DualSPHysics implementation can be found in the official documentation and related literature~\cite{crespo2015dualsphysics,wang20163d,fourtakas2016modelling,dominguez2022dualsphysics}.

In this work, our multi-agent system is developed around the non-Newtonian module in DualSPHysics v5.0. Specifically, the generalized Herschel--Bulkley--Papanastasiou (HBP) constitutive model is adopted to represent the rheology of complex materials (e.g., slurries, pastes, and suspensions), which is appropriate for debris flow simulation. The constitutive relation is written as
\[
\boldsymbol{\tau} = \eta_{app} \dot{\boldsymbol{\gamma}}
\]
where $\boldsymbol{\tau}$ is the shear stress tensor, $\dot{\boldsymbol{\gamma}}$ is the shear rate tensor, and $\eta_{app}$ is the apparent viscosity. The shear-rate tensor is computed from the velocity gradient tensor as
\[
\dot{\boldsymbol{\gamma}} = \nabla \boldsymbol{v} + (\nabla \boldsymbol{v})^{T}.
\]
and the apparent viscosity is formulated as
\[
\eta_{app} = \mu \|\dot{\boldsymbol{\gamma}}\|^{n-1} + \frac{\tau_y}{\|\dot{\boldsymbol{\gamma}}\|}\left(1 - e^{-m\|\dot{\boldsymbol{\gamma}}\|}\right).
\]
Here $\mu$ is the viscosity (or consistency index), $n$ is the power-law index (commonly referred to as the Herschel--Bulkley parameter), $\tau_y$ is the yield stress of the fluid, and $m$ is the exponential shear-rate growth parameter (or Papanastasiou parameter). The magnitude of the symmetric shear-rate tensor is defined as

\[
\left|\dot{\boldsymbol{\gamma}}\right|=\sqrt{\frac{1}{2}\left((\operatorname{tr}(\dot{\boldsymbol{\gamma}}))^{2}-\operatorname{tr}(\dot{\boldsymbol{\gamma}}^{2})\right)}
\]
By tuning the parameters $\mu$, $n$, $m$, and $\tau_y$, the model can represent a variety of non-Newtonian flow behaviors, including shear thinning, shear thickening, as well as the limiting cases of Newtonian and Bingham plastic fluids. In DualSPHysics, the HBP rheological model contributes to the deviatoric stress term in the SPH formulation of the linear momentum equation, while the pressure--density relation defines the pressure term. Together with the continuity equation, these governing equations are advanced through an explicit time-integration scheme to update particle density, stress, velocity, and position.

Without an AI-agent, a typical manual DualSPHysics v5.0 workflow is as follows: the user first prepares an .xml file that specifies geometry, physical parameters (e.g., viscosity $\mu$ and density $\rho$), numerical parameters (e.g., speed of sound $c_s$ and artificial viscosity $\alpha_{\mu}$ for numerical stabilization), and run controls (e.g., total physical simulation time and output interval). After the .xml file is prepared, a pre-processing script calls the binary executable \texttt{GenCase} to generate the initial particle configuration. Data in .vtk format are also produced so the user can visually inspect the initial setup in tools such as ParaView. The user then launches the main \texttt{DualSPHysics} executable, either in CPU mode or with GPU acceleration. Simulation outputs are saved in \texttt{.bi4} format at the selected time interval, and a post-processing script can convert \texttt{.bi4} files to \texttt{.vtk} or \texttt{.csv} that stores the computed attributes for each particle pending further analyses. These pre-processing, simulation, and post-processing steps can also be chained into a single automated run.

\subsection{Agentic System Design Principles: Agentic Workflow or Agent?}
A central challenge in applying LLMs to computational simulation is deciding where control flow should reside. Two paradigms have emerged~\cite{anthropic2024agents, microsoft2025workflows}:

\begin{enumerate}
  \item \textbf{Agent:} An LLM dynamically determines which steps to take based on context and available tools. Instead of predefining the execution path, the model reasons about what to do next at each turn. This can take the form of a single AI-agent with access to all tools, or a multi-agent system in which specialized agents coordinate through inter-agent communication.
  \item \textbf{Agentic Workflow:} A predefined sequence of operations in which the execution path is explicitly defined in code. AI-agents can be embedded as components at specific nodes, but the overall flow is controlled programmatically.

\end{enumerate}
The key distinction lies in control over the execution path: AI-agents fully delegate runtime decision-making to the LLM, while agentic workflows encode the execution sequence explicitly in programmatic logic.

To determine which paradigm fits DualSPHysics, we examine the
existing human-driven engineering workflow. A simulation engineer starts by conducting pre-processing configurations. Then a fixed sequence of executables
(\texttt{GenCase} $\rightarrow$ \texttt{DualSPHysics}
$\rightarrow$ post-processing tools) is run, each consuming the
output of the previous stage. Finally, the engineer performs
analysis tailored to the investigation.

The simulation execution stage is \emph{deterministic}: given
valid input, the tool invocation sequence is fixed and requires
no adaptive decision-making. Flexibility, and therefore the need for LLM
reasoning, is concentrated at the boundaries: upstream in input
generation and downstream in post-processing.

This structure exposes the limitations of pure agent-based approaches. Whether single- or multi-agent, the LLM must rediscover the correct execution sequence through reasoning at every run, allowing unnecessary, and potentially erroneous decisions about steps that should execute unconditionally. A multi-agent system further introduces coordination overhead for stages that require no intelligence. While either variant might arrive at the correct behavior, neither guarantees it, and both consume tokens on steps where no reasoning is needed.

We therefore adopt an agentic workflow architecture with LLM-powered AI-agents embedded only at the nodes where adaptive reasoning adds genuine value. The simulation pipeline executes programmatically, while AI-agents handle input generation and post-processing, which are the stages where flexibility is actually required.

\subsection{Three Phase Agentic Workflow with Human in the Loop}
\begin{figure*}[t]
    \centering
    \includegraphics[width=0.8\textwidth]{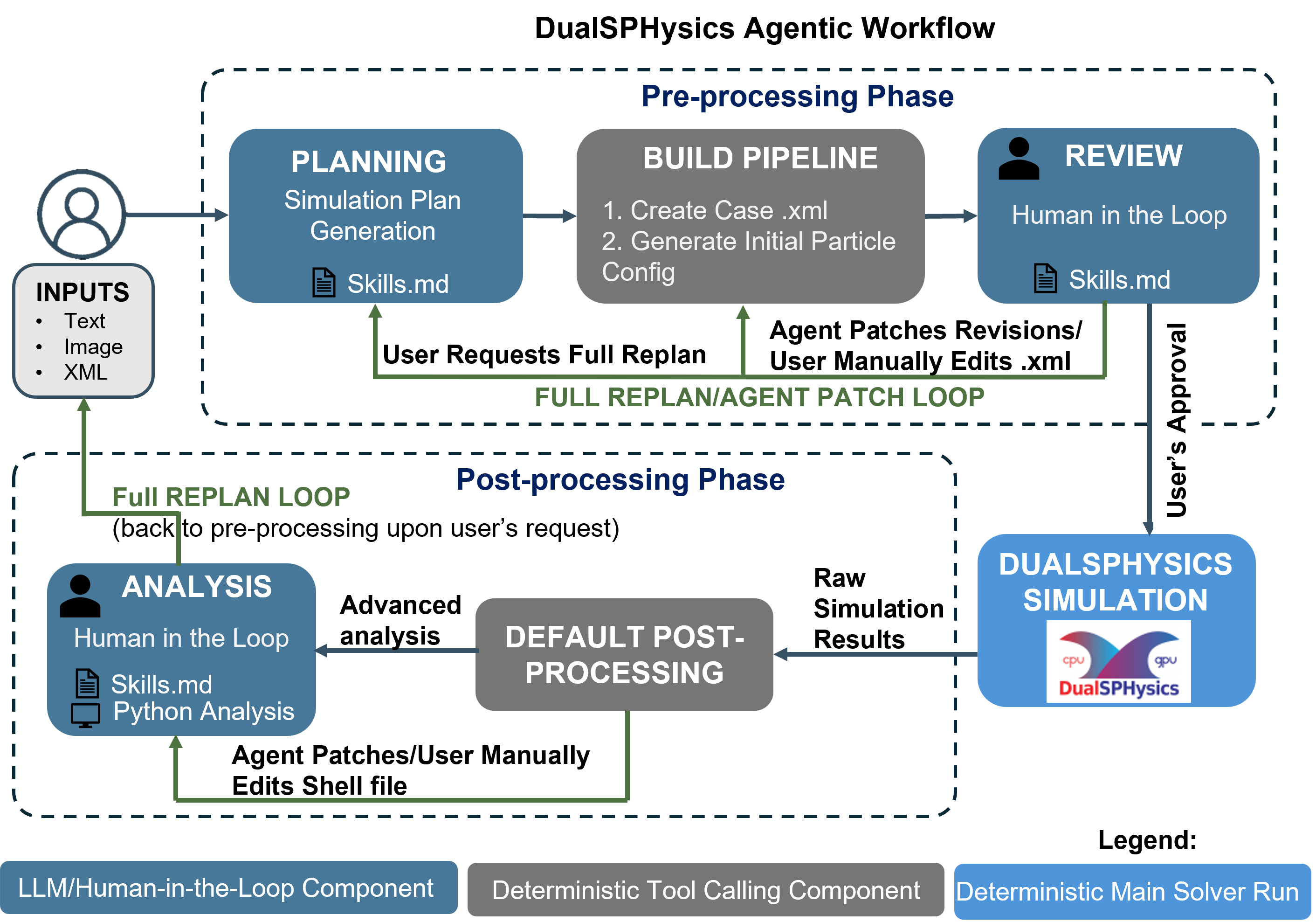}
    \caption{Three-phase agentic workflow with-human-in-the-loop configuration.}
    \label{fig:agent-structure}
\end{figure*}

The system is organized as a three-phase agentic workflow with human in the loop, as shown in Figure~\ref{fig:agent-structure}.

\begin{enumerate}
\item \textbf {Phase 1: Pre-processing} 
In pre-processing, the user provides natural language, an image, or a combination of both to define the initial geometry and other settings. If natural language or an image is provided, a planning AI-agent, which is equipped with a skill set derived from the DualSPHysics documentation and example code, will produce a structured simulation plan in an .xml file. Here, XML (Extensible Markup Language) is a hierarchical, tag-based format used to encode simulation geometry, physical and numerical parameters, and run controls in a machine-readable and reproducible manner. The workflow also accepts direct .xml input and revises on top of it. A deterministically built pipeline then executes three steps in fixed order: (1) convert the user commands into a base .xml file, (2) call \texttt{GenCase} executable file to produce the initial particle configuration, and (3) render a visualization for user inspection. Once the pipeline run is completed, the system enters a human-in-the-loop stage: the generated plan, visualization, and .xml file are presented to the user who can freely converse with the AI-agent. This includes asking questions about the setup, requesting changes or start over, or allowing the user to directly edit the .xml file. The simulation proceeds only when the user explicitly grants the approval.

\item \textbf {Phase 2: Simulation} 
The main SPH simulation is conducted in this phase. The agentic workflow launches DualSPHysics main solver. GPU availability is detected automatically. This phase is fully deterministic and requires no user interaction.

\item \textbf {Phase 3: Post-processing} 
Once the main simulation is completed, a default post-processing shell script is triggered that calls a default post-processing executable file to convert raw \texttt{.bi4} data into readable .vtk format. The user can then perform an initial inspection of the visual results. An interactive post-processing stage is subsequently initiated, in which the user may request additional ad-hoc analyses, such as extracting particle trajectories, identifying fluid free-surface profiles, or computing reaction forces on boundary walls. To fulfill these requests, the AI-agent determines an appropriate execution path based on the available tools. The AI-agent may invoke additional, non-default DualSPHysics post-processing binaries by generating and executing corresponding shell scripts. If the built-in tools prove insufficient or ill-suited for the task, the AI agent will instead generate and execute custom Python scripts, either as standalone solutions or to process the data generated by those executable files. Similar to Phase 1, the AI-agent is equipped with a knowledge base skill file derived from DualSPHysics documentation and example cases. The user may interactively refine the workflow by modifying generated scripts through natural language instructions or direct editing. This stage also supports iteration: based on post-processing insights, the workflow can return to Phase 1 for revision of the simulation setup.
\end{enumerate}

\subsection{Multimodal Input}

The system supports three input modes for specifying a simulation scenario:
\begin{enumerate}
\item \textbf {Text} The user describes the scenario in natural language. For example, simulate a debris flow in a 4 m channel with density 1500 kg/m³.
\item \textbf {Image} The user provides an image or sketch of the desired geometry. The image is sent to the AI-agent in the pre-processing phase as a multimodal input alongside the text description.
\item \textbf {XML} The user directly provides a DualSPHysics format case .xml file. The system matches the user's description to available files and injects the matched .xml into the planning prompt as context.
\end{enumerate}

Empirically, image input produces better results than text-only descriptions for complex geometries. Describing spatial relationships, dimensions, and boundary configurations in natural language is inherently ambiguous and often requires verbose, cumbersome specifications. A labeled diagram, by contrast, conveys the same information unambiguously, allowing vision-language models like GPT 5.4 to extract precise dimensions, shapes, and spatial relationships directly. We provide a detailed comparison of image-based and text-only inputs in Section~\ref{subsec:geometry-eval}.

\section{Evaluations}
\begin{figure*}[t]
    \centering
    \includegraphics[width=1.0\textwidth]{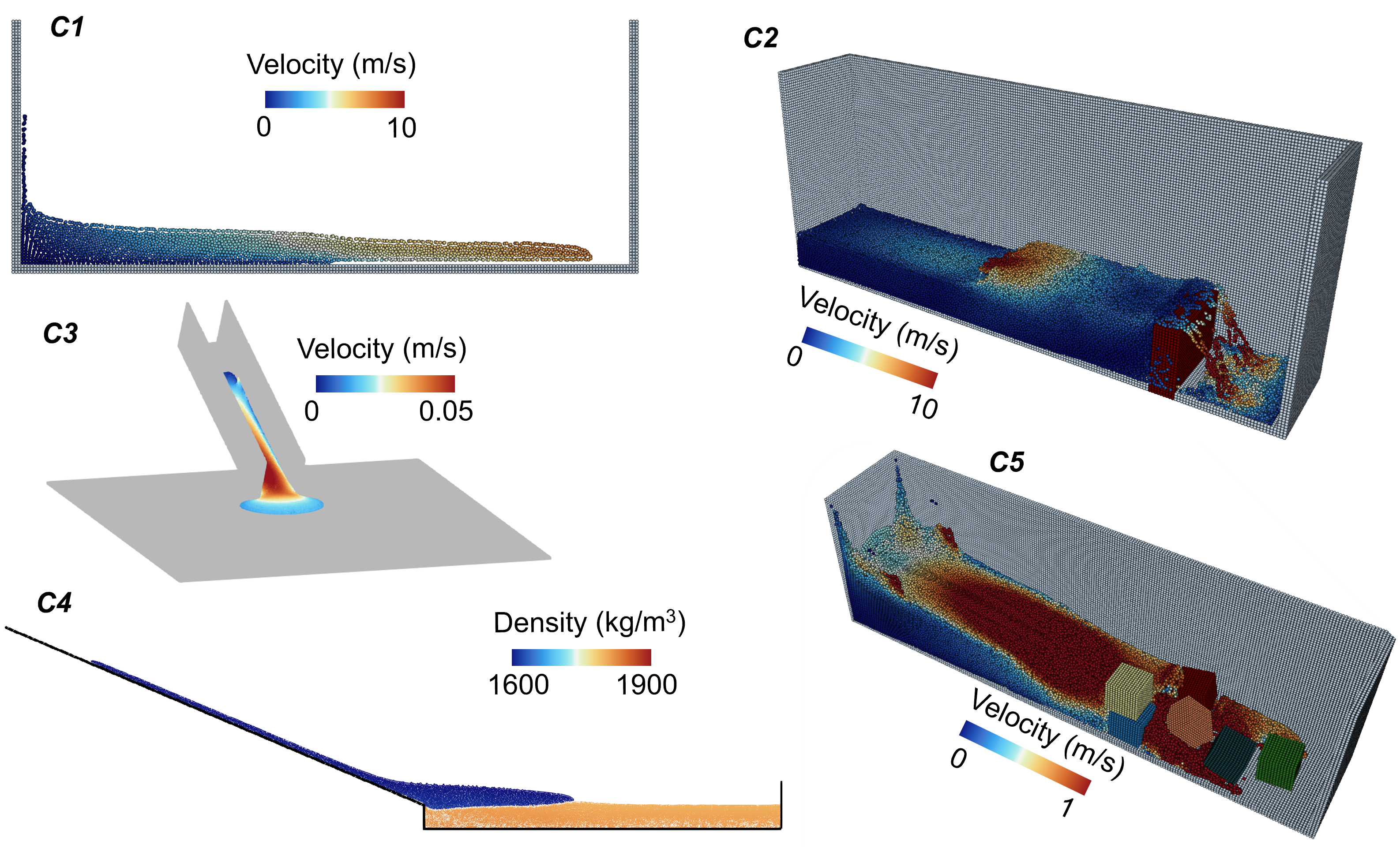}
    \caption{Summary of the five evaluation cases (C1--C5) used to assess the proposed agentic workflow across different geometries, physics, and analysis tasks.}
    \label{fig:case-summary}
\end{figure*}

We evaluate the proposed agentic workflow using five representative cases (C1--C5), with each case repeated at least three runs (R1--R3). Figure~\ref{fig:case-summary} presents snapshots of the final states of these five cases. Together, they are designed to span across various geometric and physics complexity, dimensionality, and analysis difficulty, thereby testing the robustness of the agentic workflow under a range of realistic debris flow scenarios. The cases are evaluated independently and the results are aggregated by cognitive task type rather than by case to enable cross-case comparison. All runs employed the same LLM model of GPT-5.4.

\textbf{Case 1} (Figure~\ref{fig:case-summary} C1) considers the collapse and run-out of an initially rectangular two-dimensional (2D) debris mass. This is the simplest benchmark case and is used to assess whether the agentic workflow can reliably understand the geometry, generate the setup, run the simulation, and track fundamental outputs such as flow shape evolution and velocity-related field quantities.

\textbf{Case 2} (Figure~\ref{fig:case-summary} C2) extends the problem to three dimensions (3D) by allowing an initially rectangular debris mass to collapse and impact a small barrier. This case evaluates whether the AI-agent can correctly interpret and construct 3D geometries and whether it can support analysis of more complex impact-related physics.

\textbf{Case 3} (Figure~\ref{fig:case-summary} C3) considers a 3D debris mass flowing from an inclined trench onto a flat deposition surface, mimicking alluvial-fan formation in nature. This case is designed to test the robustness of the agentic workflow for irregular geometries and to examine whether it can transfer knowledge from published studies~\cite{pandey2025integrating} to perform analogous post-processing tasks in a new but related setting.

\textbf{Case 4} (Figure~\ref{fig:case-summary} C4) examines a 2D debris mass flowing from an inclined plane onto another debris layer. This case is used to evaluate the AI-agent's ability to set up multiphase or multi-body flow configurations and to analyze deformation and interaction between the different material phases.

\textbf{Case 5} (Figure~\ref{fig:case-summary} C5) is similar to Case 2, except that the fixed barrier is replaced by six movable blocks that can be displaced by the collapsing debris. This configuration tests the AI-agent's ability to pre-process floating or mobile objects and to track their subsequent motion during the simulation.

\subsection{Pre-Process Geometry Setup Performance}
\label{subsec:geometry-eval}
\begin{figure*}[t]
\centering
\includegraphics[width=1.0\textwidth]{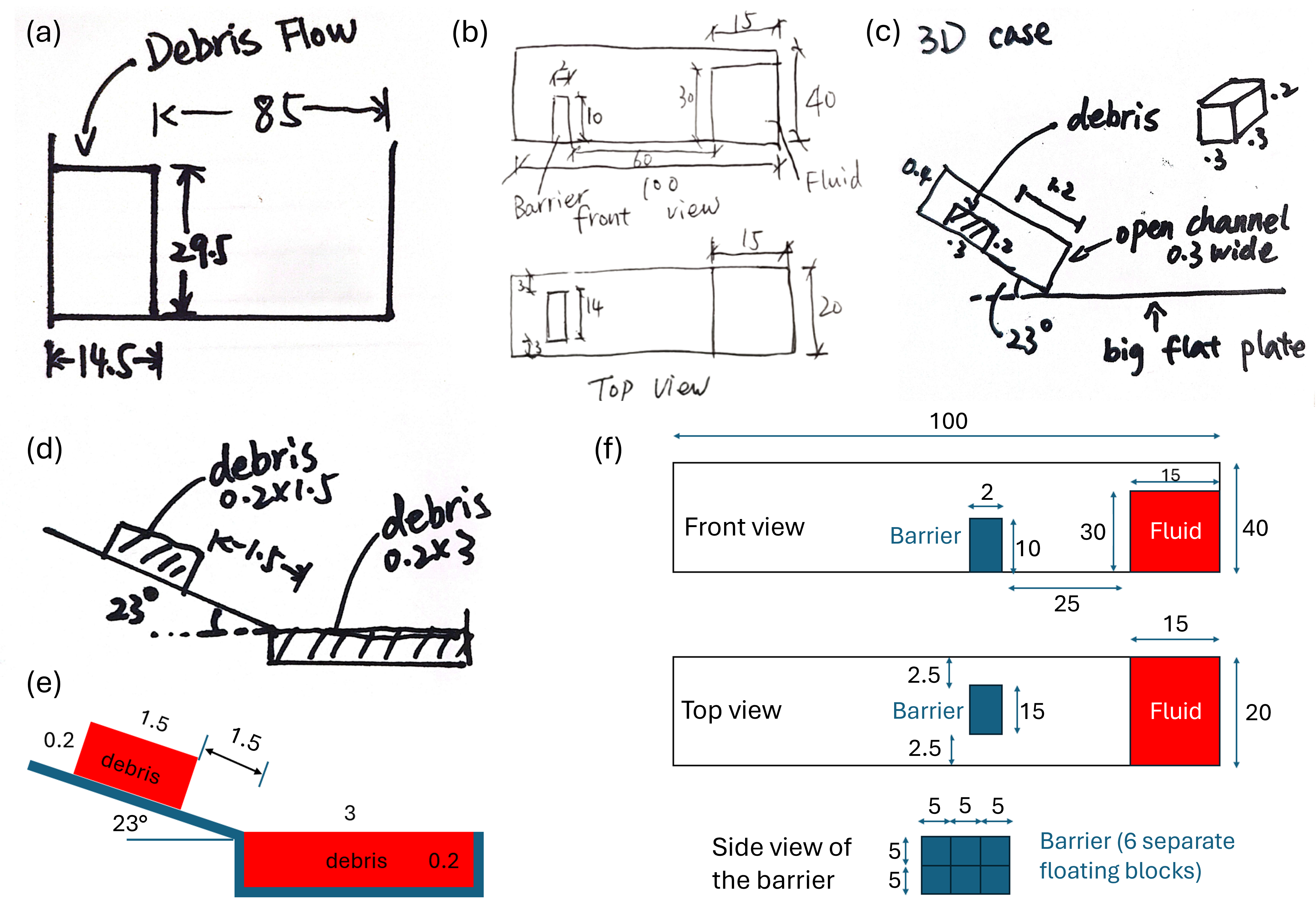}
\caption{Sketches of pre-processing geometries. Hand-drawing sketches for cases (a) C1, (b) C2, (c) C3, (d) C4, and sketches using PowerPoint for cases (e) C4*, (f) C5. All length units are meter.}
\label{Figure_PreProcessing}
\end{figure*}

\begin{figure*}[t]
\centering
\includegraphics[width=1.0\textwidth]{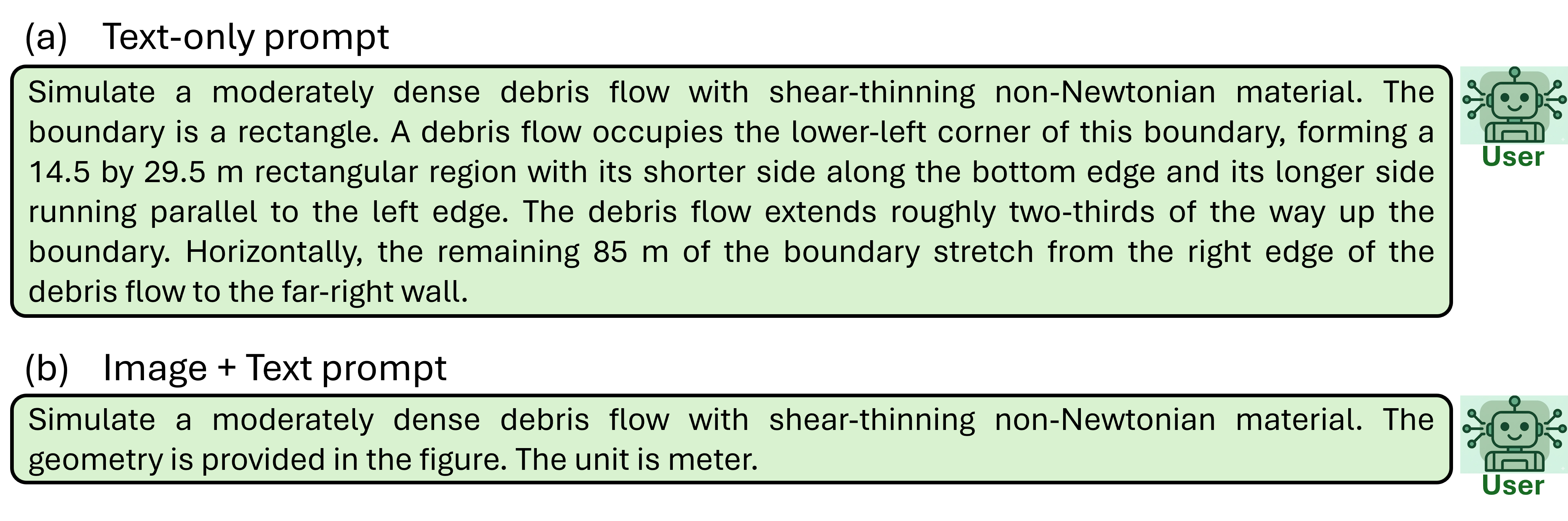}
\caption{Differences of user input prompts for (a) text-only and (b) image + text (see Figure~\ref{Figure_PreProcessing}(a) for the image prompt). Exampled by case C1.}
\label{Fig_Image_Text_Prompt}
\end{figure*}

We evaluated pre-processing performance across Case 1 to Case 5 with varied geometric complexity. The ability of the agentic workflow to interpret setup geometry was assessed using the sketches shown in Figure~\ref{Figure_PreProcessing}. For Cases C1-C4, the inputs consist of hand-drawn sketches (Figure~\ref{Figure_PreProcessing}(a)-(d)). Hand sketching reflects how engineers naturally communicate geometry in practice, and supporting this modality is essential for practical usability. Case C4* and Case C5 instead use higher-fidelity images created in PowerPoint (Figure~\ref{Figure_PreProcessing}(e)-(f)), allowing us to examine whether improved visual quality enhances geometric interpretation. For image-based inputs, a concise accompanying text prompt specifies only the physical settings of each case, without restating geometric details. This modality is referred to as \textit{image + text prompt}. For comparison, all cases are also evaluated using a \textit{text-only prompt}, where both geometry and physical parameters must be described purely in language. Figure~\ref{Fig_Image_Text_Prompt} illustrates representative differences between these two modalities. Note other input constraints like debris flow material properties are provided in subsequent dialogues that are not included in Figure~\ref{Fig_Image_Text_Prompt}.

Two metrics are used to evaluate geometry generation. The zero-shot pass rate is defined as the fraction of runs in which the generated geometry satisfies all correctness criteria on the first attempt. The human-in-the-loop (HITL) count records the number of iterative correction turns required before the geometry is accepted; runs that fail to converge within five interaction rounds are classified as failures. Table~\ref{tab:results} summarizes both metrics across all cases.

\begin{table*}[t] 
\centering
\caption{Geometry evaluation results across all cases.} 
\label{tab:results}
\vspace{0.5em}
\begin{tabular}{lcccc}
\toprule
 \rowcolor{gray!15}
 & \multicolumn{2}{c}{\textbf{Text-Only}} 
 & \multicolumn{2}{c}{\textbf{Image+Text}} \\
\cmidrule(lr){2-3} \cmidrule(lr){4-5}
 \rowcolor{gray!15}
 Case & Zero-Shot & HITL Rounds 
 & Zero-Shot & HITL Rounds \\
  \rowcolor{gray!15}
 & Pass Rate  & (avg.)      
 & Pass Rate  & (avg.) \\
\midrule
C1  & 0/3 & 1.33 & 2/3  & 0.33 \\
C2  & 0/3 & 1.33 & 0/3 & 1.33  \\
C3  & 0/3 & $\geq$5 & 0/3 & $\geq$5 \\
C4  & 0/3 & $\geq$5 & 0/3 & $\geq$5 \\
C4* & \multicolumn{2}{c}{-} & 0/3 & 4 \\
C5  & 0/3 & 1.67 & 0/3 & 2.33 \\
\bottomrule
\end{tabular}
\end{table*}

Across both modalities, zero-shot pass rates are near zero, indicating that geometry generation is not yet reliable as a fully autonomous step. The task requires coordinating multiple interdependent constraints, including dimensions, spatial placement, overlap avoidance, and boundary definitions, which are difficult to fully specify in an initial prompt. Even carefully constructed inputs tend to omit implicit assumptions that only become apparent after visualizing the generated geometry. This limitation is particularly pronounced in the text-only modality, where spatial relationships must be inferred purely from language. Consequently, a human-in-the-loop verification stage remains necessary to identify and correct errors before they propagate into downstream simulation.

An exception is Case C1 under the image + text modality, which achieves a relatively high zero-shot pass rate (2/3) and a low average HITL count (0.33). This case corresponds to the simplest geometry in the benchmark, suggesting that visual input is effective when the geometric configuration is unambiguous and low in complexity. However, this advantage does not persist as complexity increases. In Case C5, the image + text modality requires more HITL iterations than the text-only modality to reach an acceptable geometry. A plausible explanation is that, in the current setup, the image carries the full geometric specification while the accompanying text omits geometric details. As a result, the AI-agent must infer all spatial relationships from the sketch alone, which becomes increasingly error-prone for complex configurations.

These results suggest that image-only geometry specification (even when paired with non-geometric text) is insufficient for complex cases. A combined representation—where geometry is redundantly and explicitly encoded in both visual and textual modalities—may improve robustness. However, this hypothesis requires further systematic validation. 

We further annotate each failed run with a failure mode drawn from a six-class taxonomy:
\begin{itemize}
    \item \textbf{F1 -- Wrong dimensionality.} The overall geometry is correct, but one or more dimensions deviate from the specified values (e.g., 80 m instead of 75 m). Note that F1 is recorded only when the discrepancy stems from incorrect geometric extents; apparent dimensional offsets caused by fluid--boundary interface misalignment or insufficient boundary thickness are attributed to F2 and F3, respectively.
    \item \textbf{F2 -- Fluid--boundary interface inconsistency.} Fluid particles either penetrate into the boundary region or leave a spurious gap from it. In both sketch-based and text-based prompts, geometric descriptions often imply that the fluid domain is bounded by walls without explicitly specifying the required particle-level separation or contact. For example, in Figure~\ref{Figure_PreProcess_errors}(a), the debris mass is described as sitting at the corner of the tank, which implicitly requires particles to be positioned close but entirely within the boundary without overlap. A robust AI-agent should infer this constraint and enforce a consistent interface. This issue is specific to particle-based methods such as SPH, where domain boundaries are represented by discrete particles and require explicit spatial arrangement. In contrast, mesh-based methods typically enforce boundary conformity implicitly through mesh connectivity and boundary condition assignment, avoiding such interface inconsistencies.
    \item \textbf{F3 -- Boundary thickness violation.} The boundary is under-resolved, typically consisting of only a single layer of particles. As illustrated in Figure~\ref{Figure_PreProcess_errors}(b), a minimum boundary thickness is required to ensure full kernel support for fluid particles near the wall. Without sufficient boundary layers, the support domain of a fluid particle becomes truncated, leading to inaccurate force evaluation and potential particle leakage through the boundary when using Dynamic Boundary Conditions in DualSPHysics~\cite{crespo2007boundary,crespo2015dualsphysics}.  This failure mode is also specific to particle-based methods such as SPH, where boundary representation relies on discrete particles rather than continuous surfaces. The minimum thickness requirement is explicitly encoded in the skill files. Therefore, violating this constraint indicates that the AI-agent either failed to follow the user command, or lacked an understanding that a full kernel support is required in this specific SPH boundary treatment algorithm.
    \item \textbf{F4 -- Coordinate transformation error.} Errors in the rigid transformation between local and global coordinate frames, including incorrect rotation (e.g., geometry defined in a local slope-aligned frame but not properly rotated into the global frame) and translation offsets due to frame misalignment. Cases C3 and C4 in Figure~\ref{fig:case-summary} explicitly require coordinate-frame reasoning: the fluid geometry must be defined in the slope-aligned local frame and then transformed into the global coordinate system, rather than being placed as an axis-aligned geometry.
    \item \textbf{F5 -- XML syntax error.} The generated \texttt{.xml} file fails to parse during \texttt{GenCase} execution, indicating that the AI-agent did not correctly follow or interpret the geometry setup specifications in the user guide.
    \item \textbf{F6 -- Structural composition error.} The set of geometric components is incorrect: required components may be missing, extraneous components may be introduced, or one or more components may have an incorrect type or topology. These errors reflect qualitative mis-specification of the geometry, as opposed to dimensional inaccuracies, which are categorized under F1. Figure~\ref{Figure_PreProcess_errors}(c) shows an example of this error. In fact, this example shows multiple co-occurring errors.
\end{itemize}

\refstepcounter{figure}
\begin{center}
\includegraphics[width=0.5\textwidth]{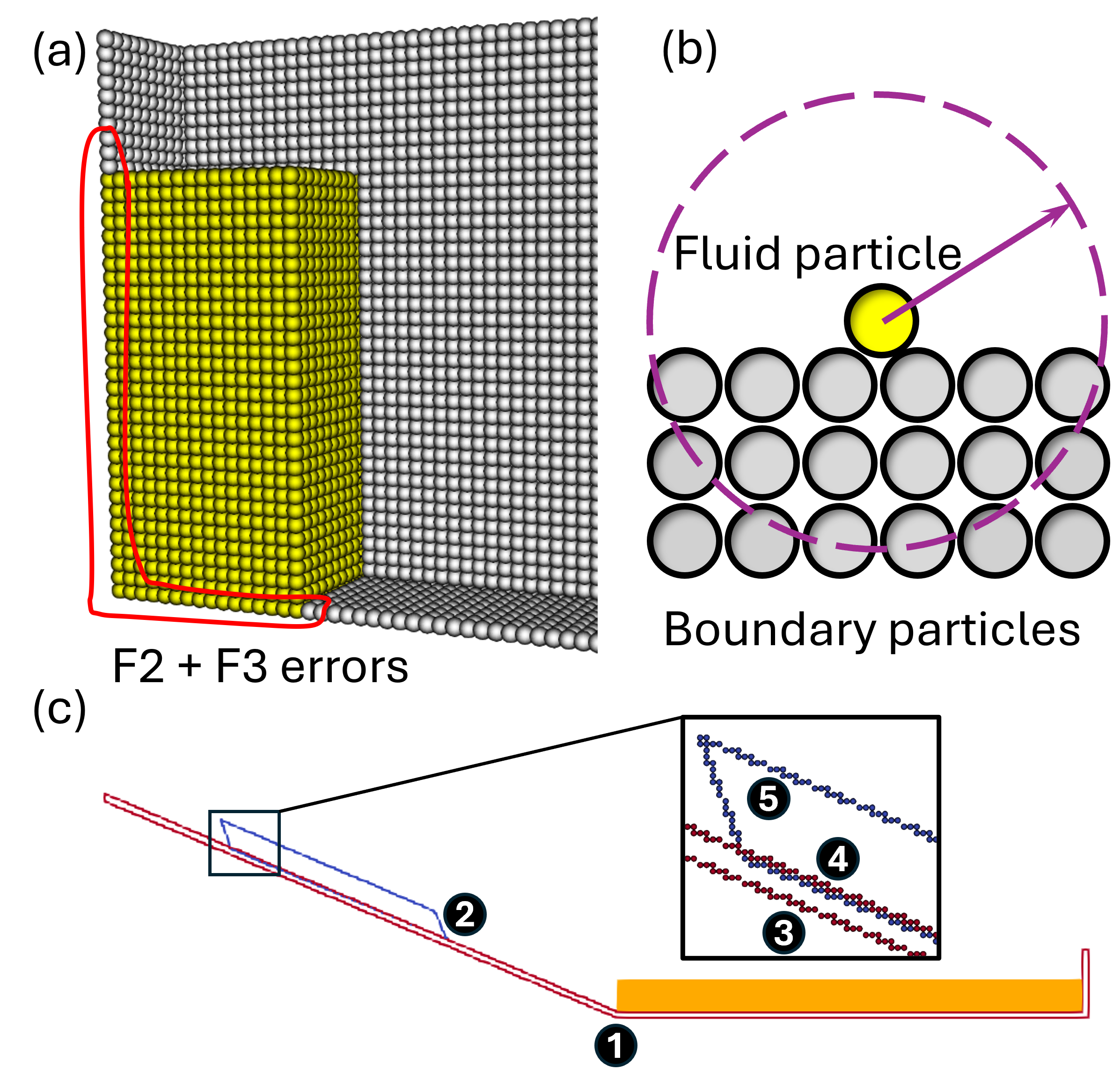}
\textbf{Figure \thefigure.} Illustration of selected zero-shot pre-processing geometry failure modes: (a) combined F2 and F3 errors in Case C2 (Run 1, text-only); (b) multiple boundary particle layers required to provide full kernel support (purple dashed region); (c) multiple failures in Case C4 (Run 3, text-only): 
\circnum{1} incorrect slope-end vertical position (F4); 
\circnum{2} debris forms a parallelogram rather than a rectangular shape (F4); 
in the zoomed-in inset:
\circnum{3} single boundary layer thickness error (F3);
\circnum{4} debris penetrates the upper boundary, indicating inconsistent rotation (F4); 
\circnum{5} missing fluid fill-in within the domain (F6).
\label{Figure_PreProcess_errors}
\end{center}

Table~\ref{tab:failure-modes} further analyzed the zero-shot failure modes for each run. For the simpler configurations (Cases C1 and C2) the AI-agent produces geometries with singular defect that can be easily corrected. Furthermore, case C2 shows that the input sketch provides only a hand-drawn front view and top view (Figure~\ref{Figure_PreProcessing}(b), yet the AI-agent reconstructs the 3D geometry without difficulty, demonstrating that it can fuse multi-view 2D input into a coherent spatial model. The dominant failure in these cases is instead the boundary thickness issue (F3) — the AI-agent frequently generates only a single boundary layer, violating the minimum-thickness constraint required for Dynamic Boundary Conditions. In fact, the hand-drawn sketches depict boundaries as single lines with no indicated thickness, which if anything understates the requirement. The failure therefore reflects a tension the AI-agent must resolve on its own: how strictly to follow skill-file requirements and judge from its own understanding of the SPH algorithm that are not reinforced — or are even visually contradicted — by the immediate prompt. These errors are nonetheless straightforward to correct, typically requiring only one or two HITL rounds, as shown in Table~\ref{tab:results}. 

\begin{table*}[t] 
\centering
\caption{Zero-shot failure modes for each run.}
\label{tab:failure-modes}
\vspace{0.5em}
\resizebox{0.96\textwidth}{!}{%
\begin{tabular}{lcccccc}
\toprule
\rowcolor{gray!15}
 & \multicolumn{3}{c}{\textbf{Text-Only}} 
 & \multicolumn{3}{c}{\textbf{Image+Text}} \\
\cmidrule(lr){2-4} \cmidrule(lr){5-7}
 \rowcolor{gray!15}
Case & Run 1 & Run 2 & Run 3 
 & Run 1 & Run 2 & Run 3 \\
\midrule
C1  & F1 & F2 & F2 
        & F1 & -- & -- \\
C2  & F2, F3 & F3 & F3 
        & F3 & F3 & F2 \\
C3  & F1,F2,F3,F4,F6 & F1,F2,F4,F6 & F1,F2,F3,F4,F6 
        & F2,F3,F4 & F5 & F5 \\
C4  & F1,F2,F3,F4,F6 & F2,F3,F4,F6 & F1,F3,F4,F6 
        & F3,F4 & F3,F4 & F3,F4,F6 \\
C4* & \multicolumn{3}{c}{-} 
        & F2,F4 & F4 & F6 \\
C5  & F3 & F3 & F2, F3 
        & F2, F3 & F3, F6 & F3, F6 \\
\bottomrule
\end{tabular}%
}
\end{table*}

Cases 3 and 4 expose a more fundamental limitation. Both configurations involve a rotated surface, which requires the AI-agent to define fluid geometry in the slope's local frame and then transform it into global coordinates. Although local-frame reasoning is documented in the skill set, the AI-agent consistently fails to apply it, with coordinate-frame errors (F4) appearing in most of the Case 3 and Case 4 run across both modalities. When the user explicitly mentions ``local coordinates'' during HITL, the AI-agent handles the slope correctly, but concurrent errors in dimensioning, placement, or boundary thickness prevent convergence within the five-round budget. We attribute part of this difficulty to the abstraction level of the hand-drawn sketches, which may underspecify features such as boundary thickness and relative proportions. To test this hypothesis, we re-ran Case 4 with a PowerPoint-generated image in which boundary thickness and geometric relationships are explicitly depicted (Case 4*). While the zero-shot pass rate remains 0/3, all three runs converge at approximately four HITL rounds rather than exceeding 5 times (Table~\ref{tab:results}), and the per-run failure modes at zero shot are fewer and less severe (Table~\ref{tab:failure-modes}). This suggests that image fidelity is a meaningful lever: abstract sketches suffice for simple geometries, but complex configurations benefit from more detailed visual input.

Case 5 presents a geometry closely related to Case 2, with the key difference that the barrier is composed of six floating blocks rather than a continuous structure. The AI-agent correctly encodes these floating blocks in the .xml file. Beyond the F3 boundary-thickness issue observed in Case 2, however, Case 5 exhibits an additional failure mode F6: the side view of the barrier, as shown in Figure~\ref{Figure_PreProcessing}(e) is always mistaken as the front view by the AI-agent. Correcting this mistake requires on average two HITL rounds. This case also demonstrates that providing only image-based geometry may be suboptimal for representing certain complex configurations. A multimodal approach combining text and imagery is believed to yield superior results, which will be investigated in future work.

Taken together, the statistical results favor text+image-based specification over text-only prompts. Geometry is difficult to describe exhaustively in prose, whereas a sketch conveys spatial relationships directly and with less user effort. The zero-shot failure modes in Table~\ref{tab:failure-modes} are in general fewer under Image+Text than under Text-Only for complex scenarios, and when the zero-shot output is closer to correct, the subsequent HITL conversation is more productive—the user corrects residual issues rather than rebuilding the geometry from scratch.

\subsection{Post-Process Analysis Performance}
Evaluating an AI-agent for HITL-ended scientific analysis tasks presents a structural challenge not present in standard benchmark settings: the difficulty of a task is co-determined by the AI-agent's capability and the specificity of the user's prompt. A task requiring four dialogue turns to resolve may reflect a genuinely capable AI-agent recovering from an ambiguous instruction, or an incapable AI-agent that could not succeed even with a fully specified one. These two scenarios carry very different implications for deployment and should not be conflated under a single "turns to completion" metric.
We therefore score every (task × run) instance along two independent dimensions. Prompt clarity (PC) characterizes the information content of the user's initial request, scored on a three-level ordinal scale: PC=1 denotes a fully specified prompt in which the reference geometry, coordinate system, physical domain, and calculation method are all stated explicitly; PC=2 denotes a partially specified prompt in which the key physical concept is stated but geometric or methodological detail is left implicit; PC=3 denotes an under specified prompt in which only the high-level concept is given and all domain interpretation is delegated to the AI-agent. AI-agent capability (AC) characterizes the AI-agent's response conditional on the prompt, scored as: A — correct on first attempt; B — correct after self-correction without additional domain knowledge from the user; C — correct only after the user supplied missing domain knowledge; F — failed despite explicit expert guidance. This decomposition allows prompt quality and AI-agent capability to be analyzed independently and permits cross-case aggregation: a result scored (PC=2, A) is directly comparable across cases regardless of geometric differences between them.

Across the five benchmark cases, post-processing requests were grouped into five cognitive task types based on the primary reasoning demand placed on the AI-agent. \textbf{Scalar / curve extraction tasks} (type 1) require reading particle data fields and computing a time-varying scalar quantity. \textbf{Visualization and rendering tasks} (type 2) require generating plots or image snapshots following explicit or implicit rendering conventions. \textbf{Group / phase identification tasks} (type 3) require partitioning the particle population into physically meaningful subsets — phases, material groups, or spatially classified regions. \textbf{Physical quantity derivation tasks} (type 4) require applying a formula (force, moment, flux) to a correctly identified particle subset. \textbf{Geometric disambiguation tasks} (type 5) require inferring the correct geometric reference from a description that leaves some ambiguity, for example, surface face, coordinate plane, cross-section orientation, or camera direction. A total of 19 tasks and 57 scored runs were evaluated across the five cases.  Examples of the full dialogue record for each of the 5 tasks can be referred in Appendix~\ref{appendix_post}.

Table~\ref{tab:full-scoring-matrix} presents the complete per-instance scoring record for all 19 tasks and 57 scored instances. Table~\ref{tab:aggregated-eval} presents aggregated evaluation results by cognitive task type, and Table~\ref{tab:pc-effect} stratifies these pass rates by prompt clarity level, separating AI-agent capability from prompt quality effects. For both Table~\ref{tab:aggregated-eval} and Table~\ref{tab:pc-effect}, pass rate is computed as A + B. Together, the three tables reveal a consistent hierarchy of AI-agent capability that cuts across case dimensionality, geometric complexity, and physics type.

Scalar extraction and visualization are robust across all cases and prompt clarity levels. Table~\ref{tab:aggregated-eval} shows a 100\% pass rate for both task types, and Table~\ref{tab:pc-effect} confirms that this robustness is largely prompt-independent: performance does not deteriorate as prompts become less specified, indicating that success in these categories reflects genuine AI-agent capability rather than favourable task framing. The three B-rated scalar extraction instances in Table~\ref{tab:full-scoring-matrix} arose because either the prompt did not fully understand the structure of the particle data in .csv file (C1-T1 Run 1) or from ambiguous command the AI-agent could not find a correct reference position in order to compute "changing of distance" (C4-T2 Runs 2 and 3). This is exampled in Fig.~\ref{Figure_example_scale}: when asking to plot the run-off distance of a debris flow, the initial and corrected run-off distance curves preserves the same shape with only shifts in space and time. Similarly, Fig.~\ref{Figure_example_viz} shows the result of the only B-rated visualization test (C2-T2 Run 2) for which the flow surface profile was requested to plot. The initial AI-agent attempt plotted both surface and interior particles rather than the debris profile (Fig.~\ref{Figure_example_viz}(a)); once the user clarified the intent, the AI-agent corrected the result without requiring further domain-specific guidance (Fig.~\ref{Figure_example_viz}(b)). These failures represent specification failures where the AI-agent's interpretation of an ambiguous prompt diverges from the user's intent. Such a failure is recoverable through more explicit task framing rather than through improvements to the AI-agent's underlying capabilities.

\end{multicols*}
\refstepcounter{table}
\begingroup
\setlength{\tabcolsep}{5pt}
\renewcommand{\arraystretch}{1.15}
\begin{center}
\small
\textbf{Table \thetable.} Full scoring matrix for 19 post-processing tasks across three run repetitions. Each cell reports prompt clarity (PC1--PC3) and AI-agent capability outcome (A/B/C/F).
\label{tab:full-scoring-matrix}
\vspace{0.5em}
\resizebox{0.98\textwidth}{!}{%
\begin{tabular}{>{\raggedright\arraybackslash}p{1.25cm}>{\raggedright\arraybackslash}p{5.6cm}>{\centering\arraybackslash}p{1.55cm}>{\centering\arraybackslash}p{2.0cm}>{\centering\arraybackslash}p{2.0cm}>{\centering\arraybackslash}p{2.0cm}}
\toprule
\rowcolor{gray!15}
\textbf{Case} & \textbf{Task Description} & \textbf{Type} & \textbf{Run 1} & \textbf{Run 2} & \textbf{Run 3} \\
\midrule
\rowcolor{gray!10}\multicolumn{6}{l}{\textbf{C1 -- 2D debris flow} \textit{(flat geometry)}} \\
C1-T1 & Run-off distance vs time & scalar & PC2 / B & PC2 / A & PC2 / A \\
C1-T2 & Wall-hit time & geodis & PC3 / C & PC3 / C & PC3 / C \\
C1-T3 & Surge height vs time & group & PC2 / A & PC2 / A & PC2 / C \\
C1-T4 & Velocity snapshot images & visual & PC2 / A & PC2 / A & PC2 / A \\
\midrule
\rowcolor{gray!10}\multicolumn{6}{l}{\textbf{C2 -- 3D debris flow + barrier} \textit{(barrier + forces)}} \\
C2-T1 & Debris front position vs time & scalar & PC1 / A & PC1 / A & PC1 / A \\
C2-T2 & Shape profile at a cross-section plane & visual & PC2 / A & PC3 / B & PC2 / A \\
C2-T3 & Barrier hit time & geodis & PC3 / C & PC2 / C & PC2 / C \\
C2-T4 & Total downstream mass & group & PC2 / A & PC3 / A & PC2 / A \\
C2-T5 & Upstream / overtop / leak \% & group & PC3 / B & PC2 / B & PC1 / A \\
C2-T6 & Particle classification viz. & visual & PC2 / A & PC2 / A & PC2 / A \\
C2-T7 & Barrier reaction force vs time & group & PC2 / C & PC2 / C & PC2 / A \\
C2-T8 & Barrier bending moments vs time & phys & PC2 / A & PC2 / A & PC2 / A \\
\midrule
\rowcolor{gray!10}\multicolumn{6}{l}{\textbf{C3 -- Inclined trench} \textit{(inclined geometry)}} \\
C3-T1 & Mass flux rate at trench exit & phys & PC3 / C & PC2 / C & PC2 / C \\
C3-T2 & Velocity map at cross-section & visual & PC2 / A & PC2 / A & PC2 / A \\
C3-T3 & Alluvial fan profile (ref. fig.) & geodis & PC3 / C & PC3 / C & PC3 / C \\
\midrule
\rowcolor{gray!10}\multicolumn{6}{l}{\textbf{C4 -- 2D erosion} \textit{(two-phase)}} \\
C4-T1 & Sinking depth vs time & scalar & PC2 / A & PC1 / A & PC2 / A \\
C4-T2 & Run-off + deepest-point time & scalar & PC2 / A & PC3 / B & PC3 / B \\
C4-T3 & Sink / bulge volume & geodis & PC2 / A & PC2 / A & PC2 / A \\
\midrule
\rowcolor{gray!10}\multicolumn{6}{l}{\textbf{C5 -- 3D floating blocks} \textit{(moving boundaries)}} \\
C5-T1 & Block CoM per block vs time & group & PC2 / A & PC2 / A & PC3 / C \\
\bottomrule
\end{tabular}%
}
\end{center}
\endgroup
\clearpage
\refstepcounter{table}
\begingroup
\setlength{\tabcolsep}{5pt}
\renewcommand{\arraystretch}{1.15}
\begin{center}
\small
\textbf{Table \thetable.} Aggregated post-processing evaluation results by cognitive task type across 57 scored instances from 19 tasks and 5 cases.
\label{tab:aggregated-eval}
\vspace{0.5em}
\resizebox{\textwidth}{!}{%
\begin{tabular}{>{\raggedright\arraybackslash}p{2.9cm}>{\centering\arraybackslash}p{0.9cm}>{\centering\arraybackslash}p{0.8cm}>{\centering\arraybackslash}p{0.8cm}>{\centering\arraybackslash}p{0.8cm}>{\centering\arraybackslash}p{0.8cm}>{\centering\arraybackslash}p{0.8cm}>{\centering\arraybackslash}p{1.0cm}>{\raggedright\arraybackslash}p{6.8cm}}
\toprule
\rowcolor{gray!15}
\textbf{Cognitive Type} & \textbf{Tasks} & \textbf{n} & \textbf{A} & \textbf{B} & \textbf{C} & \textbf{F} & \textbf{Pass} & \textbf{Primary Failure Mode} \\
\midrule
Scalar / curve extraction & 4 & 12 & 9 & 3 & -- & -- & 100\% & CSV data structure misread; misunderstand reference position due to vague command. \\
Visualization \& rendering & 4 & 12 & 11 & 1 & -- & -- & 100\% & Strong when rendering intent is explicit; one B-ranking due to plotting the entire domain rather than just profile; self-corrected when emphasizing "profile". \\
Group / phase identification & 5 & 15 & 9 & 2 & 4 & -- & 73\% & Unable to define correct groups from ambiguous language; Defaults to static snapshot partitioning instead of tracking particles across time; read wrong data table column. \\
Physical quantity derivation & 2 & 6 & 3 & -- & 3 & -- & 50\% & Formula is usually correct, but selected wrong particle group for calculations; particle mass is interpreted incorrect. \\
Geometric disambiguation & 4 & 12 & 3 & -- & 9 & -- & 25\% & poor understanding of boundary thickness and particle-wall distance; partially failed to transfer the analysis methods from literature. \\
\bottomrule
\end{tabular}%
}
\end{center}
\endgroup
\begin{multicols*}{2}

\refstepcounter{figure}
\begin{center}
\includegraphics[width=0.95\columnwidth]{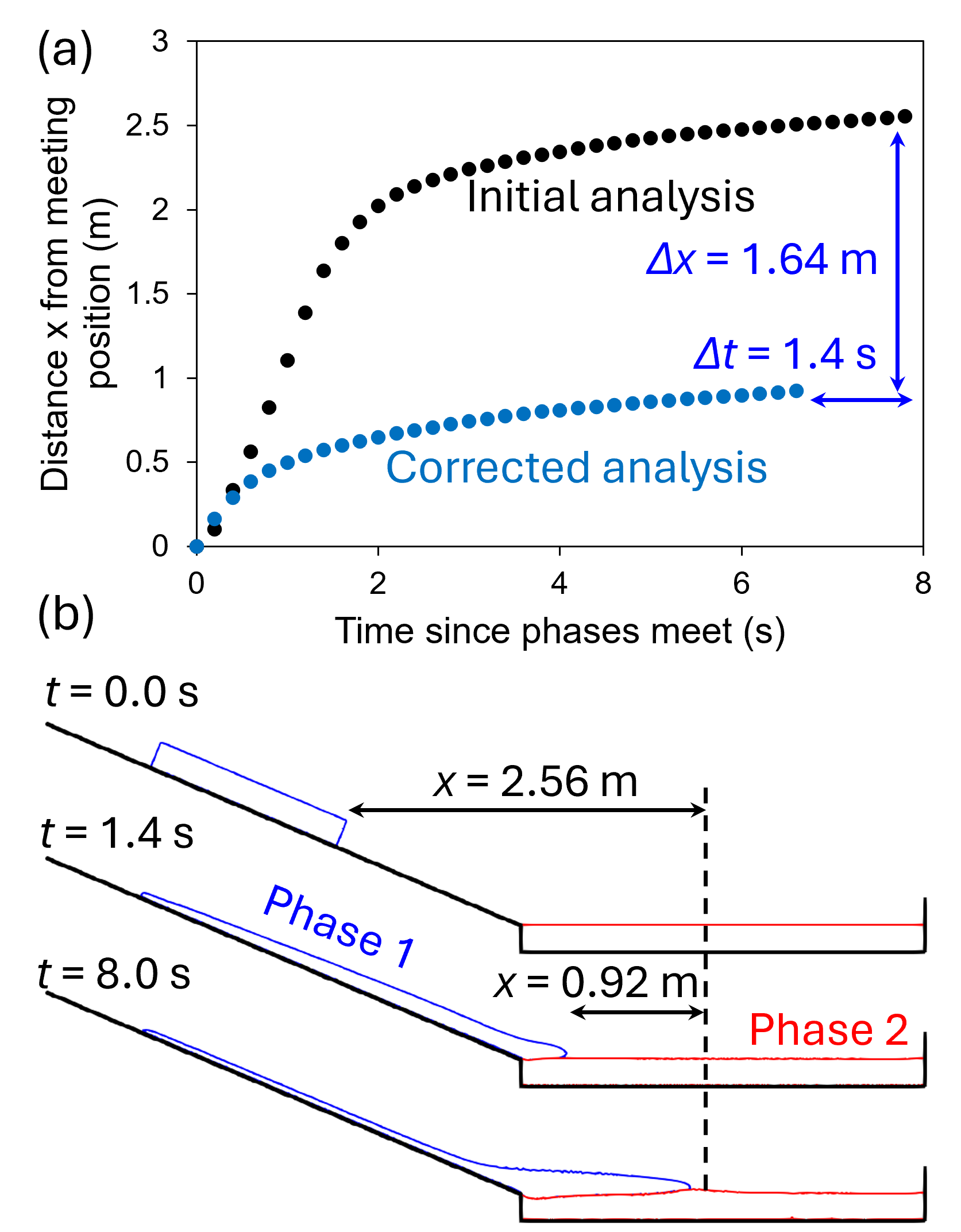}\label{Figure_example_scale}
\textbf{Figure \thefigure.} An example of AI-agent post-processing performance on the scalar-extraction task (C4-T2, Run 2) -- Run-off distance of the top debris phase starting from when the two phases meet: (a) Run-off distance versus time; (b) Illustration of the phase profiles at different time snaps.
\end{center}

\refstepcounter{figure}
\begin{center}
\includegraphics[width=0.95\columnwidth]{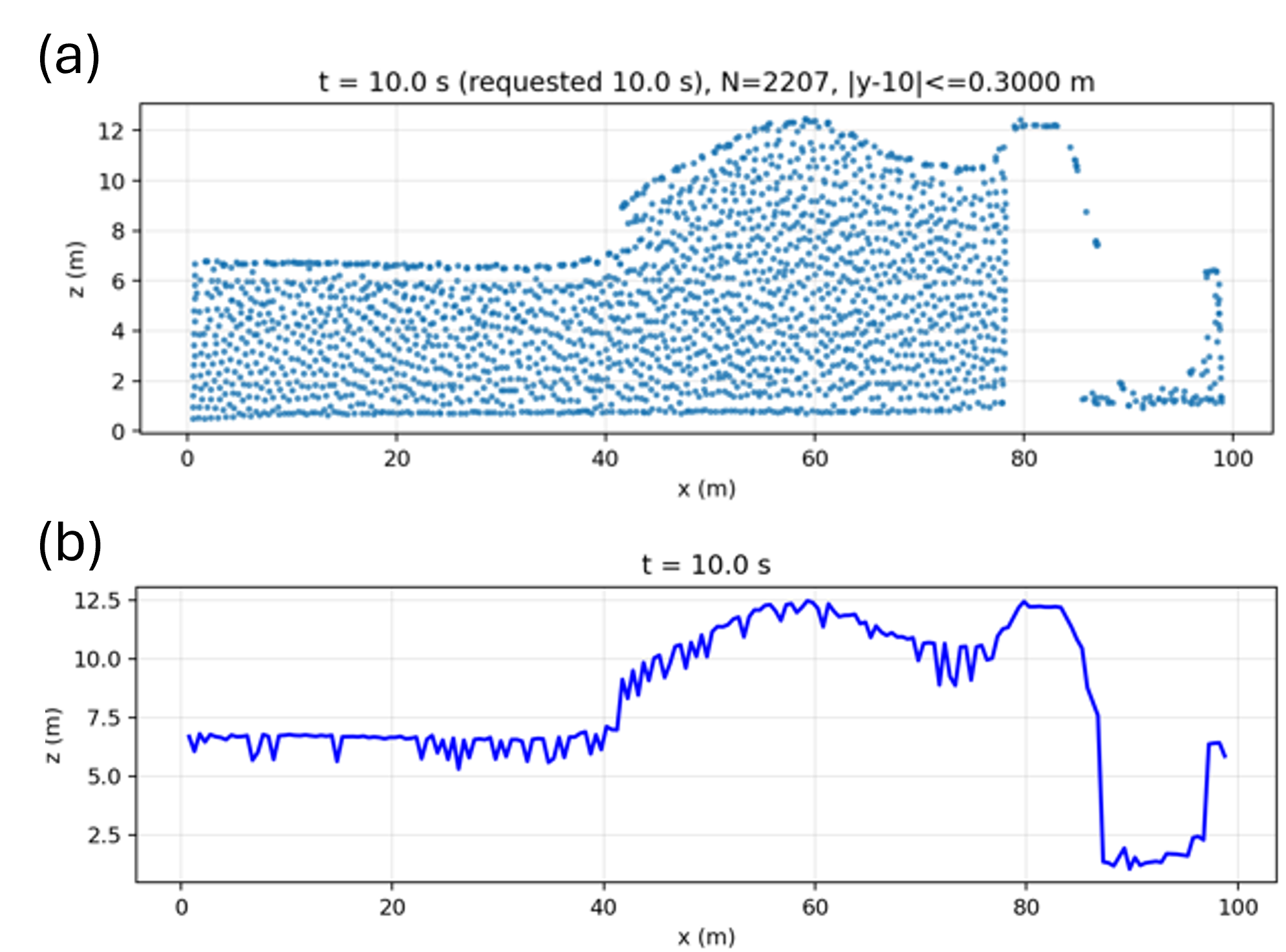}\label{Figure_example_viz}
\textbf{Figure \thefigure.} An example of AI-agent post-processing performance on the visualization and rendering task (C2-T2, Run 2) -- plotting of surface profile at plane y = 0 and t = 10 s: (a) initial plot includes all particles near the cross-section; (b) corrected surface profile plot.
\end{center}

Group and phase identification achieved an overall pass rate of 73\% (Table~\ref{tab:aggregated-eval}). Table~\ref{tab:pc-effect} shows a mild prompt-sensitivity trend — 73\% at PC$\leq$2 and 67\% at PC=3 — but given the small number of PC=3 instances ($n = 3$), this difference should not be over-interpreted. Failure modes vary by sub-task rather than by prompt clarity. For example, in C1-T3 Run 3, when asked to determine the surge height of debris after impact against the right boundary, the AI-agent failed to define an appropriate debris region from which to extract the highest point, although it succeeded in the other two repetitions. Fig.~\ref{Figure_example_group} shows another example of C2-T5 Run 1, in which the AI-agent was asked to partition the particles into three regions: those who are blocked by the barrier and remained the upstream direction, or overtopped the barrier toward the downstream direction, or leaked to the downstream side through the gap in between the barrier and the boundary. The AI-agent initially gave an erroneous answer by overestimating the leaked partition compared to the overtopped. However, it recognized that its first attempt relied only on the static particle distribution at the final time step and proposed a trajectory-based analysis. Once that proposal was approved, it produced satisfactory results. In C2-T7 Runs 1 and 2, when asked to compute the reaction force on the barrier, the AI-agent understood the force formula correctly but selected the wrong particle group, yielding a near-zero result. Correct identification of the barrier particle group in Run 3 produced a good result on first attempt. Overall, group and phase identification failures fall into two sub-categories: cases where the AI-agent correctly diagnoses its own methodological error and self-corrects (rated B), and cases where data-interpretation mistakes or ambiguous region definitions require expert intervention (rated C). The former are specification failures; the latter reflect a more persistent reasoning vulnerability in particle subset selection that also appears in physical derivation tasks.

\refstepcounter{figure}
\begin{center}
\includegraphics[width=0.95\columnwidth]{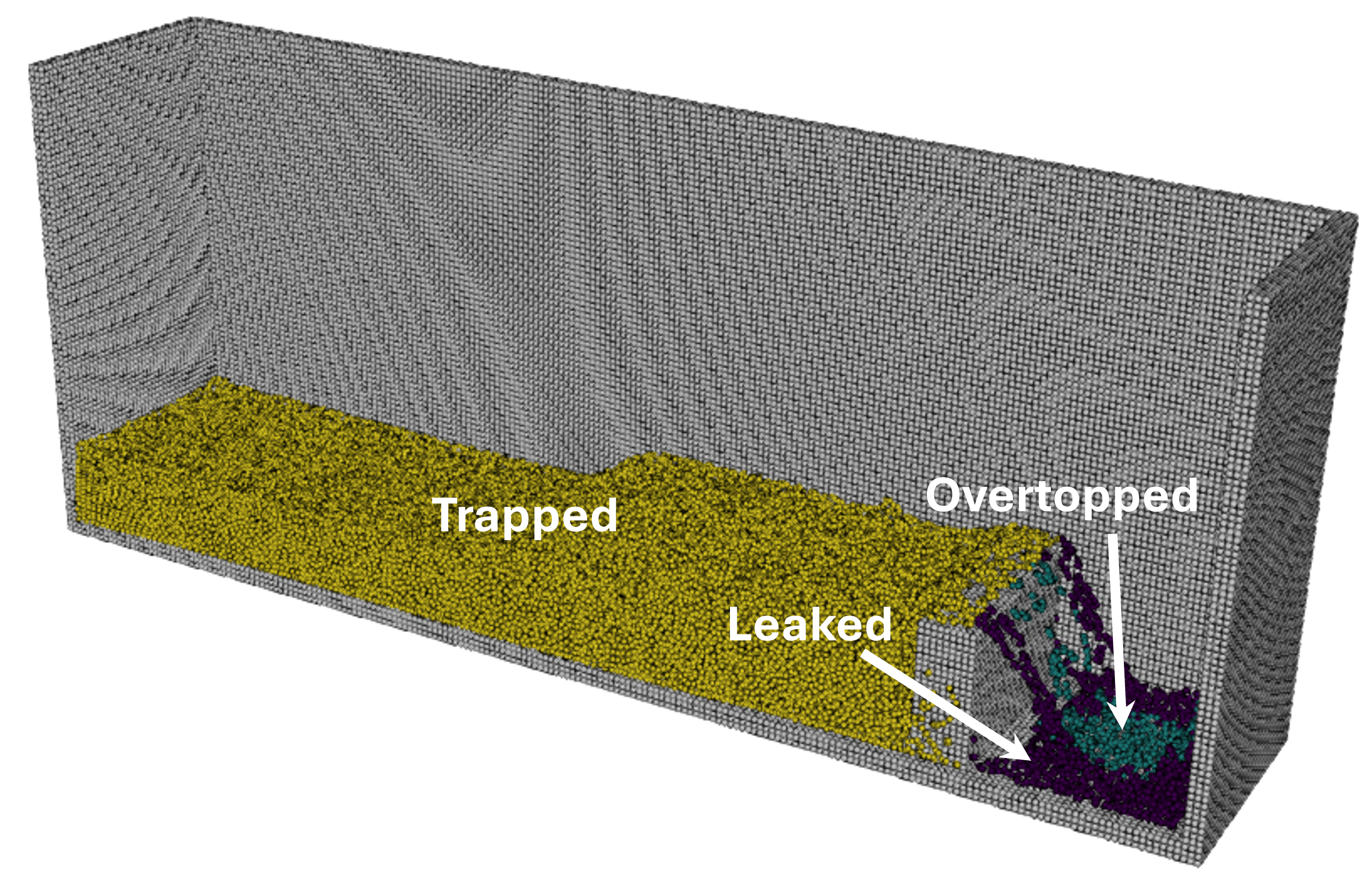}\label{Figure_example_group}
\textbf{Figure \thefigure.} An example of AI-agent post-processing performance on the group and phase identification task (C2-T6, Run 1) -- three phases of debris mass are identified: trapped by the barrier, leaked through the side of the barrier and overtopped the barrier.
\end{center}

Physical quantity derivation achieved a 50\% overall pass rate (Table~\ref{tab:aggregated-eval}). Bending moment computation (C2-T8) achieved 100\% across all three runs, benefiting from geometric context established in the preceding force task within the same session. In Fig.~\ref{Figure_example_phy}(a) (C3-T1), the AI-agent was asked to plot mass flux rate at the imaginary cross-section perpendicular to the inclined trench. This  achieved a 0\% pass rate: the AI-agent was not able to realize that each SPH particle carries a constant mass and thus, can be used to determine mass flux rate. Additionally, it also failed to correctly locate the flux cross-section at the trench exit without explicit user guidance, and it also did not understand how to read particle positions from .csv file for mass flux rate computation. Only when these pitfalls are explicitly pointed by the user the AI-agent could produce consistent and correct mass flux rate among the three repeated runs (Fig.~\ref{Figure_example_phy}(b)).  The result suggests that physical derivation performance is highly complex: the AI-agent needs to be capable of jointly identify the correct spatial reference from an implicit description, identify particle groups correctly and processed data structure correctly and understand the theoretical foundation of particle-based simulation method to solve a physical quantity derivation problem. 

\refstepcounter{figure}
\begin{center}
\includegraphics[width=0.95\columnwidth]{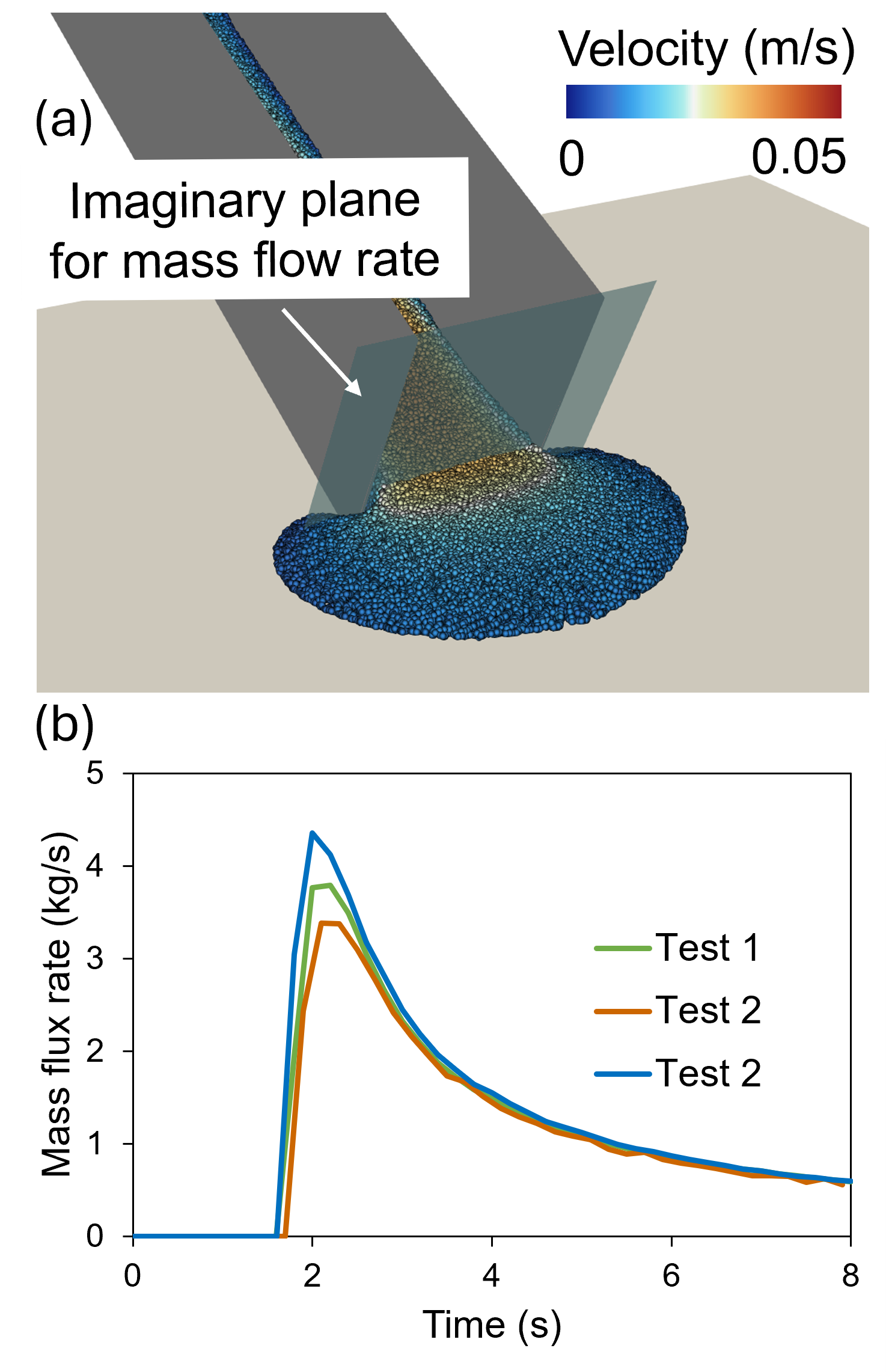}\label{Figure_example_phy}
\textbf{Figure \thefigure.} An example of the AI-agent post-processing performance on the physical quantity derivation task (C3-T1, Run 3) -- debris mass flow rate at the cross-section at the end of the trench: (a) Illustration of the plane for mass flux measurement; (b) Mass flux rate across 3 runs shows consistent result.
\end{center}

Geometric disambiguation achieved an overall pass rate of 25\% (Table~\ref{tab:aggregated-eval}) and is the weakest post-processing task type in the benchmark. Table~\ref{tab:pc-effect} shows a pass rate of 60\% at PC=2 and 0\% at PC=3. The failure modes are consistent across cases. In cases C1-T2 and C2-T3, the AI-agent repeatedly failed to determine which face of a finite-thickness boundary structure should be used for analysis. Rather than inferring the correct wall surface from the output particle positions, it tended to rely too directly to the input geometry specification and did not interpret the .xml boundary description in a physically meaningful way. As shown in Fig.~\ref{Figure_example_geo}, a related difficulty was the collision criterion: the AI-agent did not reliably recognize that, in SPH, mechanical interaction occurs over a finite kernel distance, so wall contact need not require a debris particle to geometrically cross the wall boundary. Instead, contact may already be established once particles come within interaction range or when the wall begins to register a pressure response. These struggles suggest that the AI-agent did not understand the interaction logic of SPH very well. On the other hand, in C3-T3, the AI-agent failed to infer the correct camera angle for reproducing the plan-view shape of a debris-flow alluvial fan from a published figure~\cite{pandey2025integrating}. An experienced reader can usually infer the viewing direction from figure context, but because it is not stated explicitly in the source, the AI-agent could not recover without guidance. 

\refstepcounter{figure}
\begin{center}
\includegraphics[width=0.95\columnwidth]{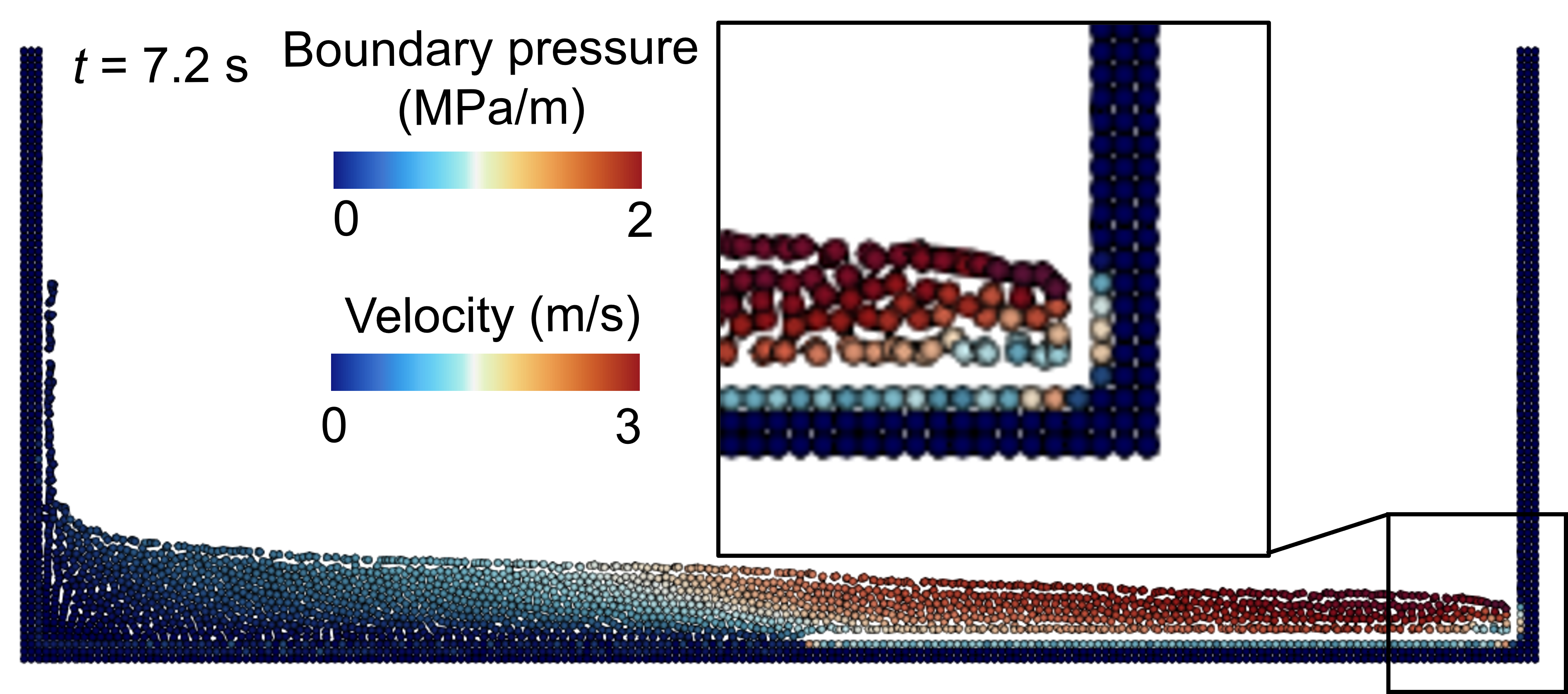}\label{Figure_example_geo}
\textbf{Figure \thefigure.} Example of AI-agent post-processing performance on the geometric disambiguation task (C1-T2-R3) at time $t = 7.2$~s, when the debris flow reaches the right boundary wall. Inset: zoomed view of the right-wall region showing the reaction pressure on the boundary particle.
\end{center}

\end{multicols*}
\refstepcounter{table}
\begingroup
\setlength{\tabcolsep}{5pt}
\renewcommand{\arraystretch}{1.15}
\begin{center}
\small
\textbf{Table \thetable.} Post-processing prompt-clarity effect on pass rate (A or B), stratified by prompt-clarity level.
\vspace{0.5em}
\resizebox{\textwidth}{!}{%
\begin{tabular}{>{\raggedright\arraybackslash}p{3.0cm}>{\centering\arraybackslash}p{1.0cm}>{\centering\arraybackslash}p{1.2cm}>{\centering\arraybackslash}p{1.0cm}>{\centering\arraybackslash}p{1.2cm}>{\centering\arraybackslash}p{1.0cm}>{\centering\arraybackslash}p{1.2cm}>{\raggedright\arraybackslash}p{6.2cm}}
\toprule
\rowcolor{gray!15}
\textbf{Cognitive Type} & \textbf{PC=1 n} & \textbf{PC=1 pass} & \textbf{PC=2 n} & \textbf{PC=2 pass} & \textbf{PC=3 n} & \textbf{PC=3 pass} & \textbf{Interpretation} \\
\midrule
Scalar / curve extraction & 4 & 100\% & 6 & 100\% & 2 & 100\% & Prompt-independent strength; perfect pass rate at all prompt-clarity levels. \\
Visualization \& rendering & -- & -- & 11 & 100\% & 1 & 100\% & Prompt-independent strength; perfect pass rate at both PC=2 and PC=3. \\
Group / phase identification & 1 & 100\% & 11 & 73\% & 3 & 67\% & Prompt-sensitive; performance degrades from PC = 1 to PC = 3. \\
Physical quantity derivation & -- & -- & 5 & 60\% & 1 & 0\% & Moderate performance under partial specification; failure under the single PC=3 run should be interpreted cautiously. \\
Geometric disambiguation & -- & -- & 5 & 60\% & 7 & 0\% & Prompt-dependent weakness; poor performance even when prompts are relatively well specified. \\
\bottomrule
\end{tabular}%
}
\label{tab:pc-effect}
\end{center}
\endgroup
\begin{multicols*}{2}

Taken together, the results across all five cognitive task types suggest two qualitatively distinct classes of failure. The first is specification failure, in which the AI-agent's interpretation of an ambiguous prompt diverges from the user's intended task definition. These failures are often recoverable through prompt refinement or lightweight clarification, and they account for all B-rated instances and a subset of the C-rated instances in the scalar and visualization categories. The second is reasoning failure, in which the AI-agent cannot infer the correct geometric reference, apply an appropriate SPH contact criterion, understand the post-processed data structure, or identify the correct particle subset for a derived quantity. These failures are much more persistent: they dominate the C-rated instances in the group and phase identification, physical quantity derivation, and geometric disambiguation categories. This distinction suggests that prompt engineering can alleviate part of the post-processing burden, but more reliable deployment will require either closer expert supervision or mechanisms that allow the AI-agent to retain and reuse corrections from previous failures. Multi-agent cooperation may also help alleviate these issues.

\section{Conclusion}
This work presents, to the best of our knowledge, the first agentic AI workflow for meshless (particle-based) simulation in computational mechanics, demonstrated on debris flow and multiphase non-Newtonian flow modeling using SPH with DualSPHysics. A key finding is the importance of human-in-the-loop interaction, particularly in pre- and post-processing. In pre-processing, multimodal inputs enable the agent to interpret geometry from sketches, and iterative user feedback effectively resolves ambiguities. Compared to text-only descriptions, integrating visual and textual inputs reduces failure modes and streamlines user communication when defining complex geometries. Furthermore, this multimodal approach mitigates several SPH-specific challenges, such as particle discretization errors and inconsistent boundary treatments. In post-processing, we introduce a cognitive-task-based evaluation framework, decomposing simulation interpretation into five fundamental tasks rather than assessing performance solely at the case level. Under this framework, the agentic workflow demonstrates strong performance in scalar and curve extraction, visualization, and phase identification, but remains less reliable in physical quantity derivation and geometric disambiguation. These results highlight the importance of structured guidance and domain-specific capabilities for further improving agent performance.

While demonstrated on SPH, the framework is general and can extend to other particle-based methods such as the Material Point Method (MPM) and Discrete Element Method (DEM), supporting applications across solid, fluid, and granular mechanics. Overall, this work highlights the potential of agentic AI to improve the accessibility and usability of physics-based simulation, enabling more effective human–AI collaboration in engineering workflows.

A future direction for increasing autonomy is the introduction of monitor agents that evaluate intermediate results and flag errors, reducing user burden while maintaining reliability.

\section{Acknowledgment}
This research was financially supported by the University of Nebraska - Lincoln.

\bibliographystyle{unsrt}
\bibliography{references}

\end{multicols*}

\appendix
\renewcommand{\thefigure}{\thesection\arabic{figure}}
\renewcommand{\theHfigure}{appendix.\thesection.\arabic{figure}}
\section{Appendix}
\setcounter{figure}{0}
\renewcommand{\thefigure}{A\arabic{figure}}
\subsection{Post-processing prompt examples}\label{appendix_post}
We present post-processing dialogue examples that covers all 5 cognitive task types. In the dialogue below, we include comments shown in the pink dialogue boxes.

\clearpage
\begin{figure}[p]
\centering
\includegraphics[width=0.95\textwidth,height=0.92\textheight,keepaspectratio]{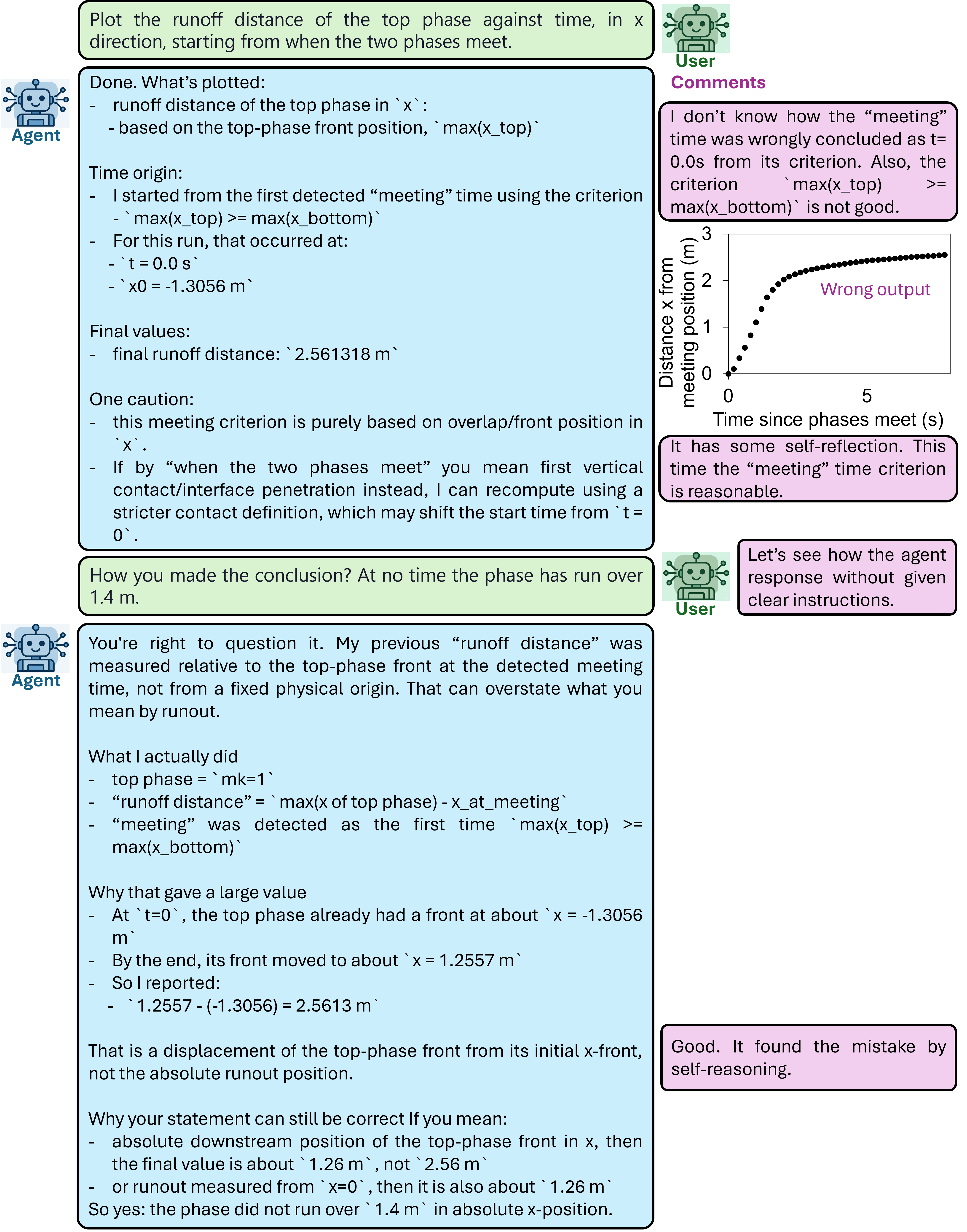}
\caption{Example post-processing dialogue for scalar / curve extraction task: Case C4, Task 2, Run 1.}
\label{fig:appendix-scale1}
\end{figure}

\begin{figure}[p]\ContinuedFloat
\centering
\includegraphics[width=0.95\textwidth,height=0.92\textheight,keepaspectratio]{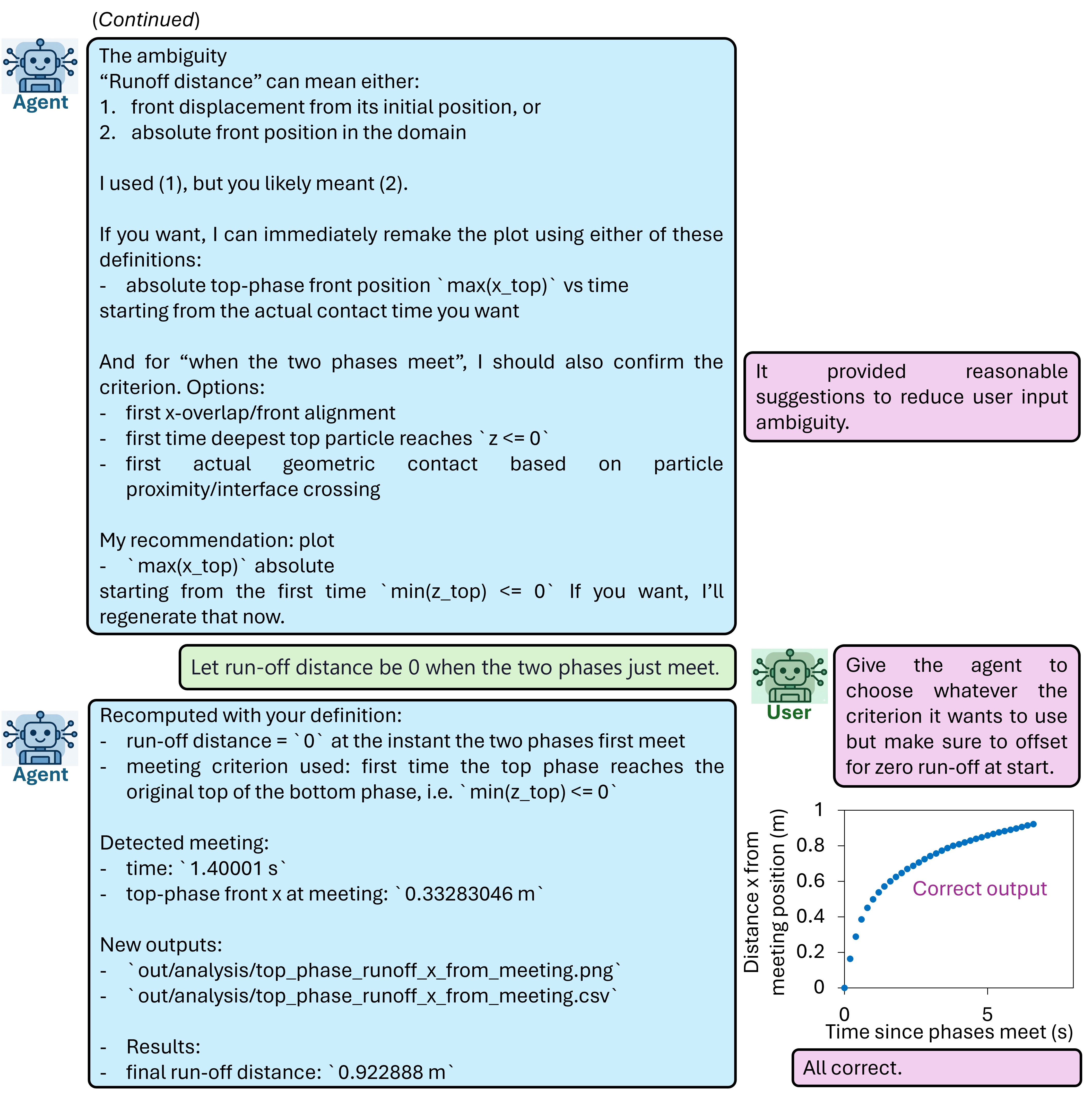}
\caption[]{(Continued)}
\label{fig:appendix-scale2}
\end{figure}

\begin{figure}[p]
\centering
\includegraphics[width=0.95\textwidth,height=0.92\textheight,keepaspectratio]{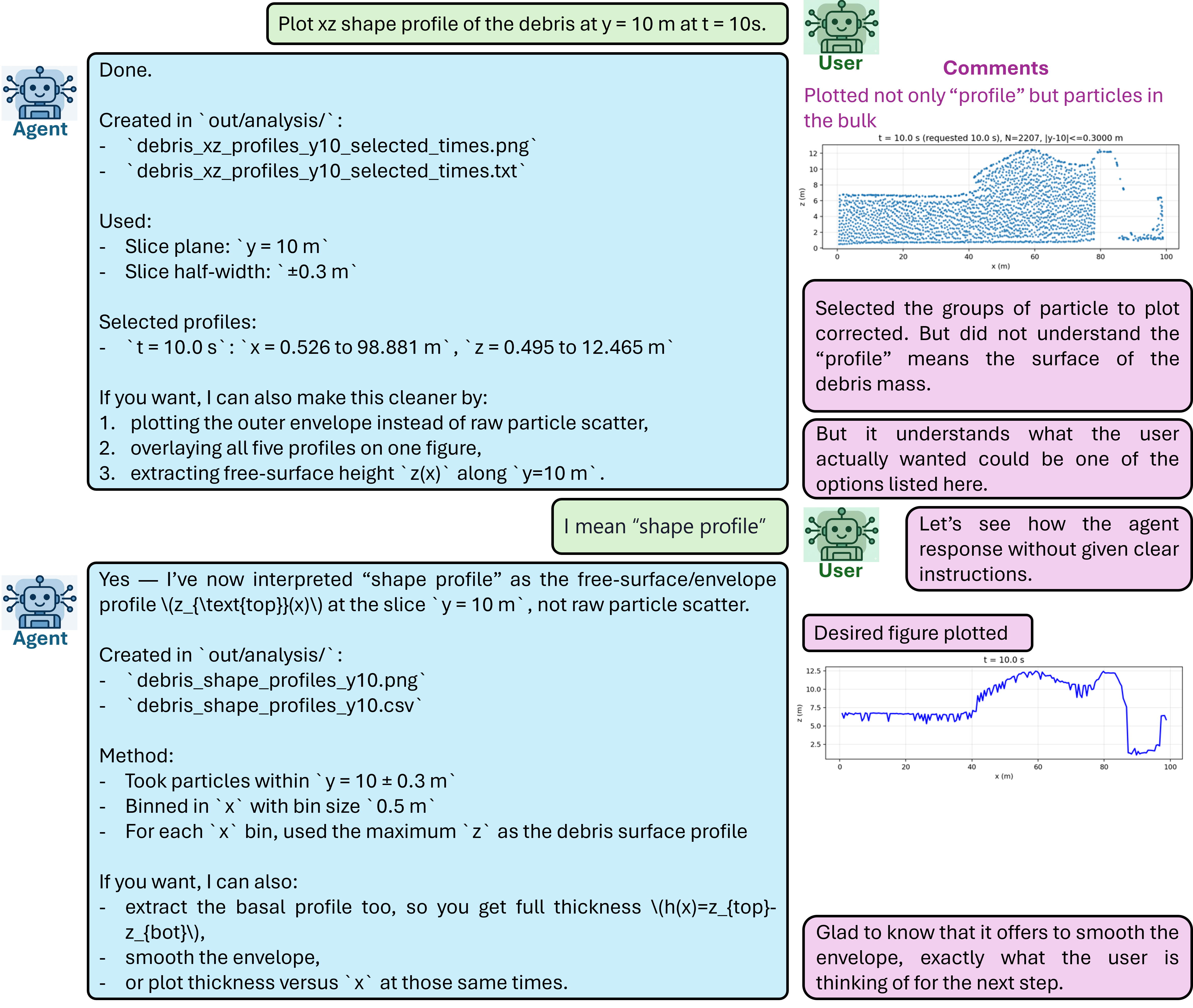}
\caption{Example post-processing dialogue for visualization \& rendering task: Case C2, Task 2, Run 2.}
\label{fig:appendix-viz}
\end{figure}

\begin{figure}[p]
\centering
\includegraphics[width=0.95\textwidth,height=0.92\textheight,keepaspectratio]{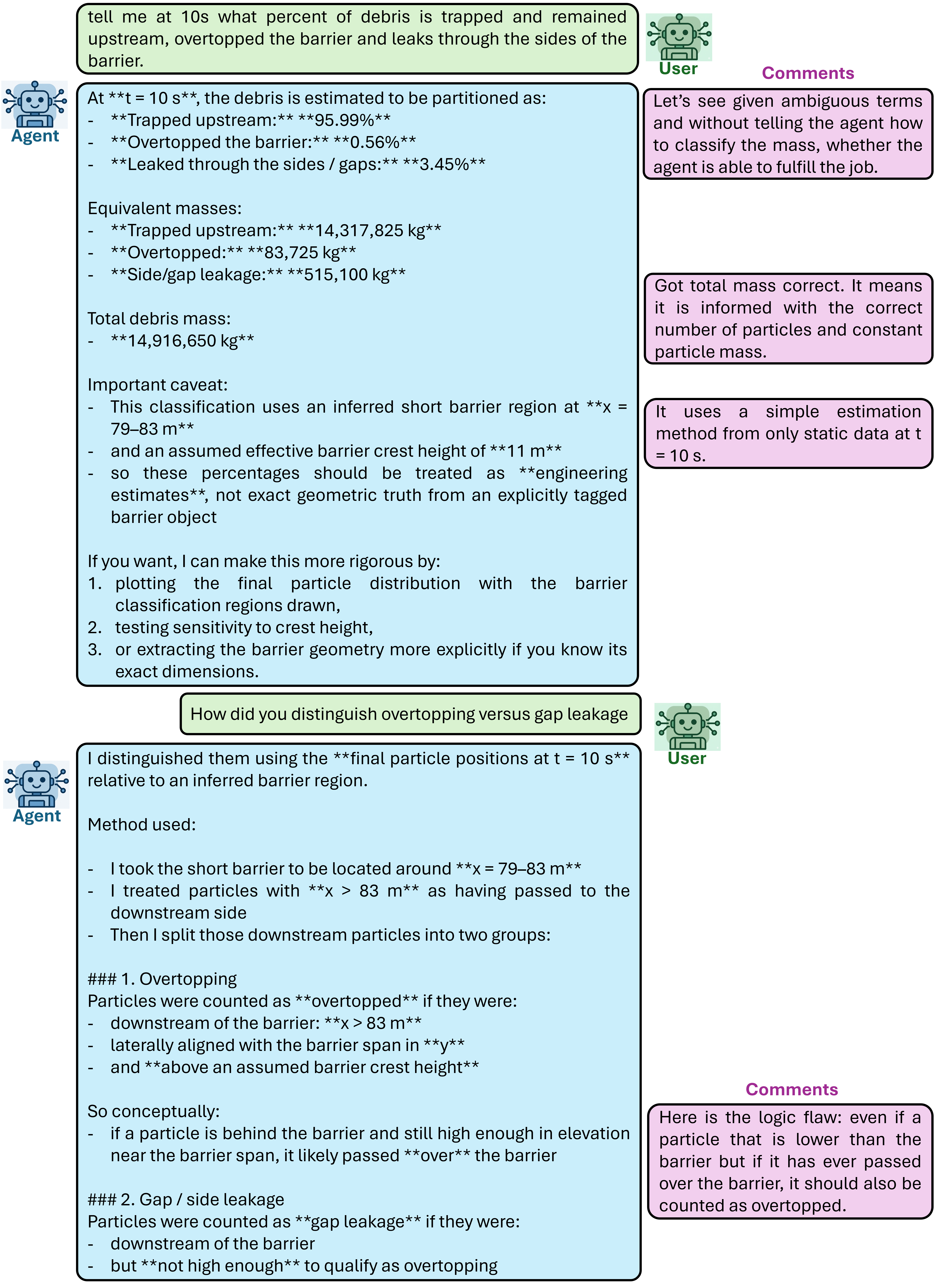}
\caption{Example post-processing dialogue for group / phase identification task: Case C2, Task 5, Run 1.}
\label{fig:appendix-group1}
\end{figure}

\begin{figure}[p]\ContinuedFloat
\centering
\includegraphics[width=0.95\textwidth,height=0.92\textheight,keepaspectratio]{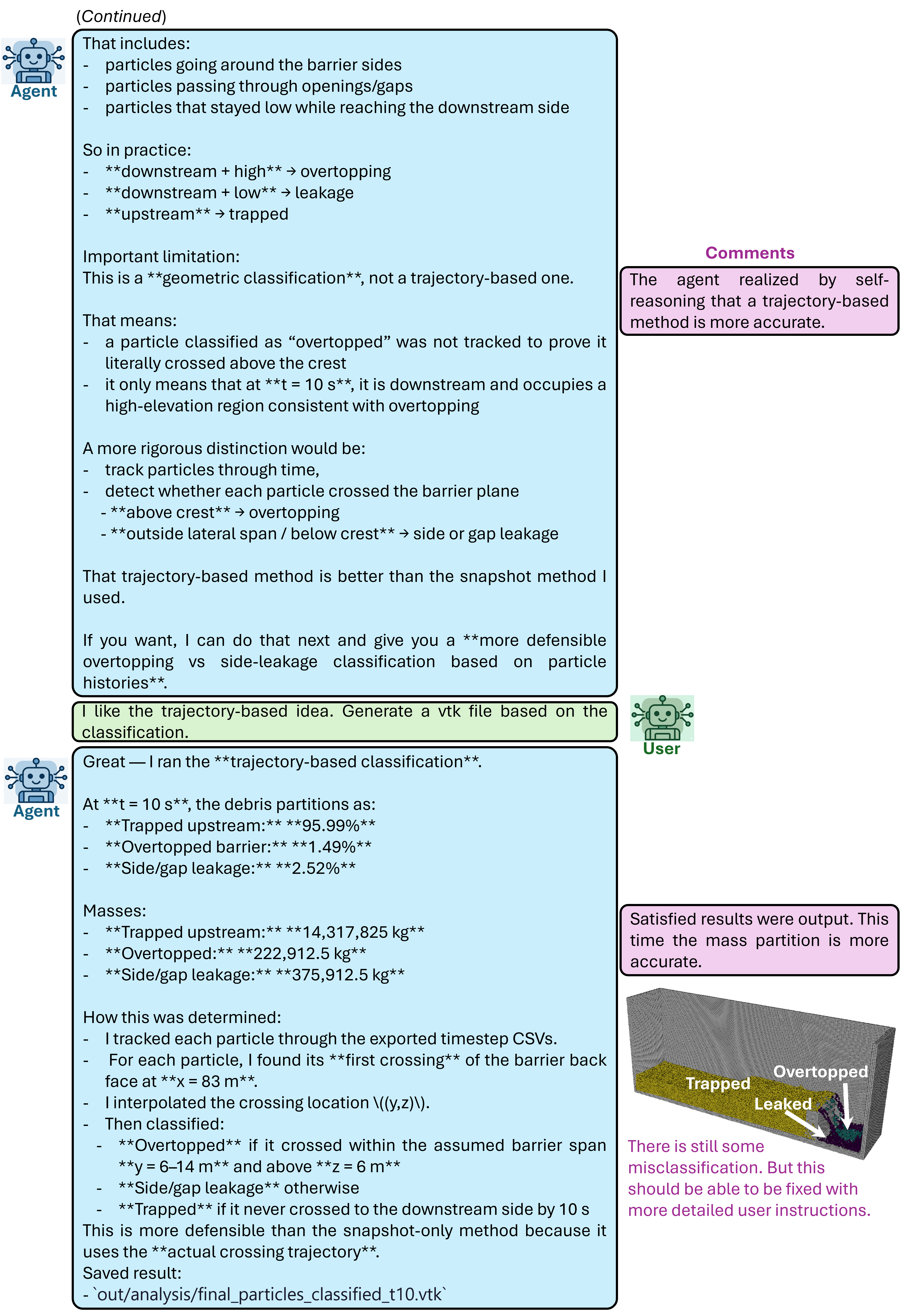}
\caption[]{(Continued)}
\label{fig:appendix-group2}
\end{figure}

\begin{figure}[p]
\centering
\includegraphics[width=0.95\textwidth,height=0.92\textheight,keepaspectratio]{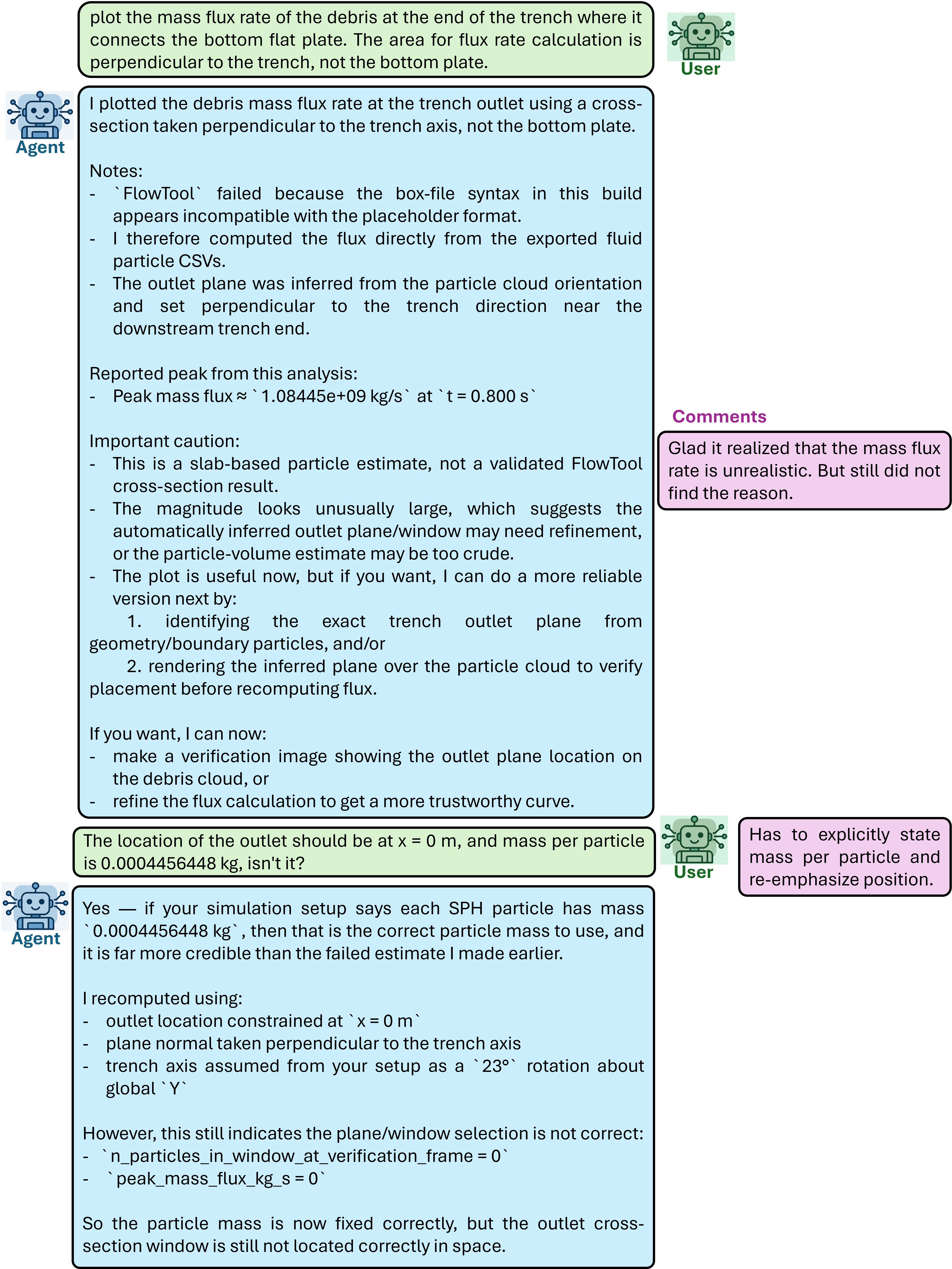}
\caption{Example post-processing dialogue for physical quantity derivation task: Case C3, Task 1, Run 2.}
\label{fig:appendix-phy1}
\end{figure}

\begin{figure}[p]\ContinuedFloat
\centering
\includegraphics[width=0.95\textwidth,height=0.92\textheight,keepaspectratio]{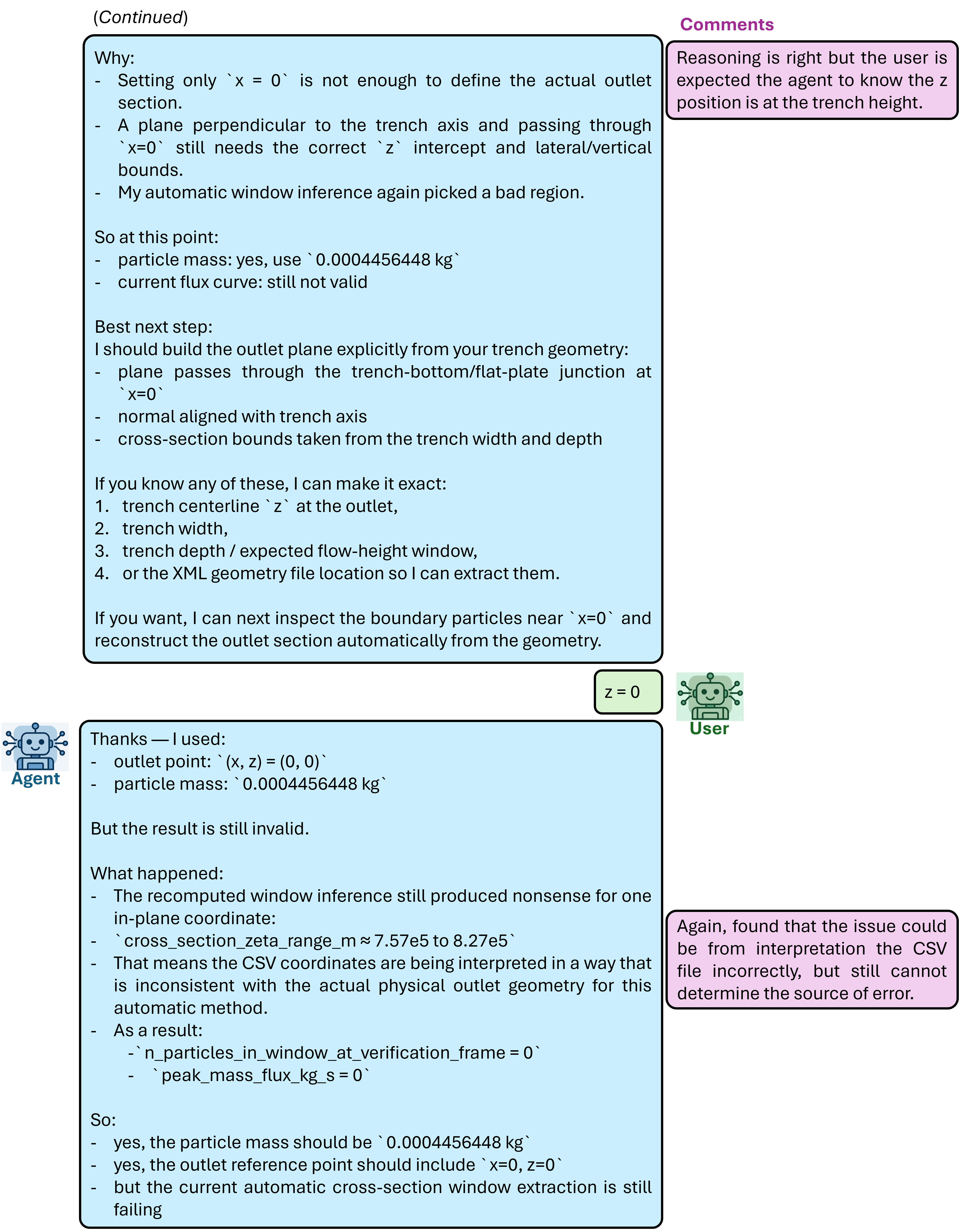}
\caption[]{(Continued)}
\label{fig:appendix-phy2}
\end{figure}

\begin{figure}[p]\ContinuedFloat
\centering
\includegraphics[width=0.95\textwidth,height=0.92\textheight,keepaspectratio]{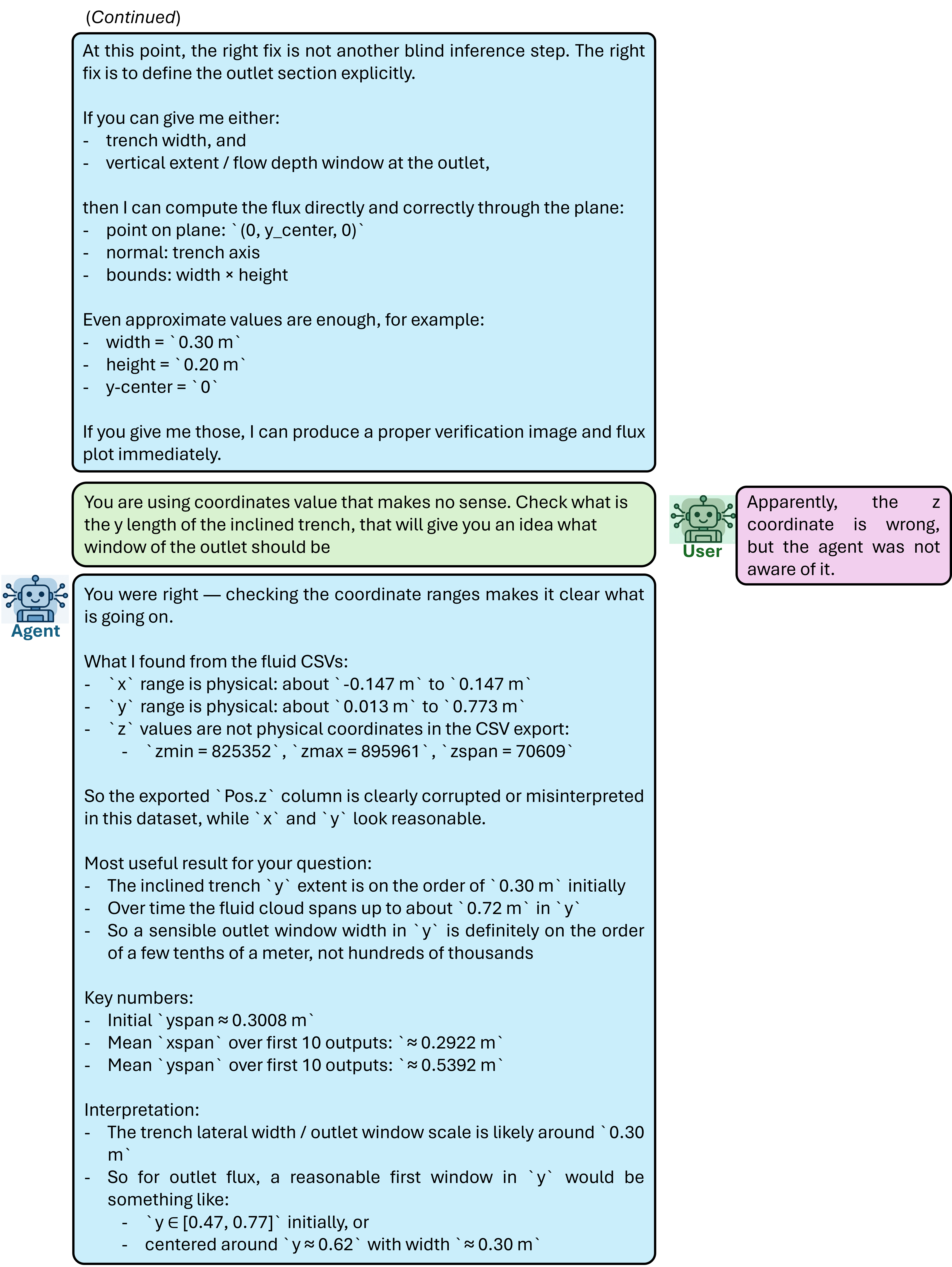}
\caption[]{(Continued)}
\label{fig:appendix-phy3}
\end{figure}

\begin{figure}[p]\ContinuedFloat
\centering
\includegraphics[width=0.95\textwidth,height=0.92\textheight,keepaspectratio]{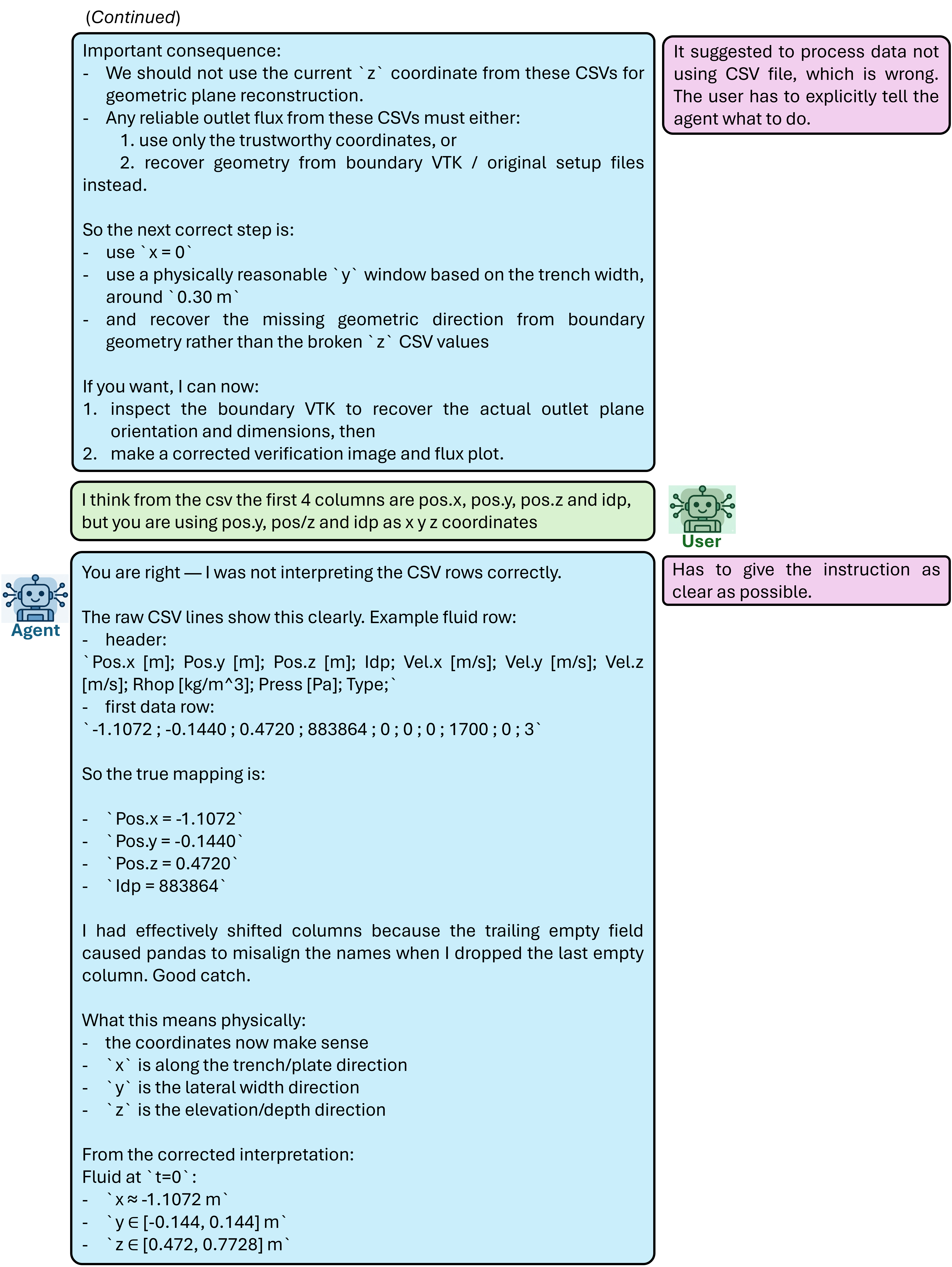}
\caption[]{(Continued)}
\label{fig:appendix-phy4}
\end{figure}

\begin{figure}[p]\ContinuedFloat
\centering
\includegraphics[width=0.95\textwidth,height=0.92\textheight,keepaspectratio]{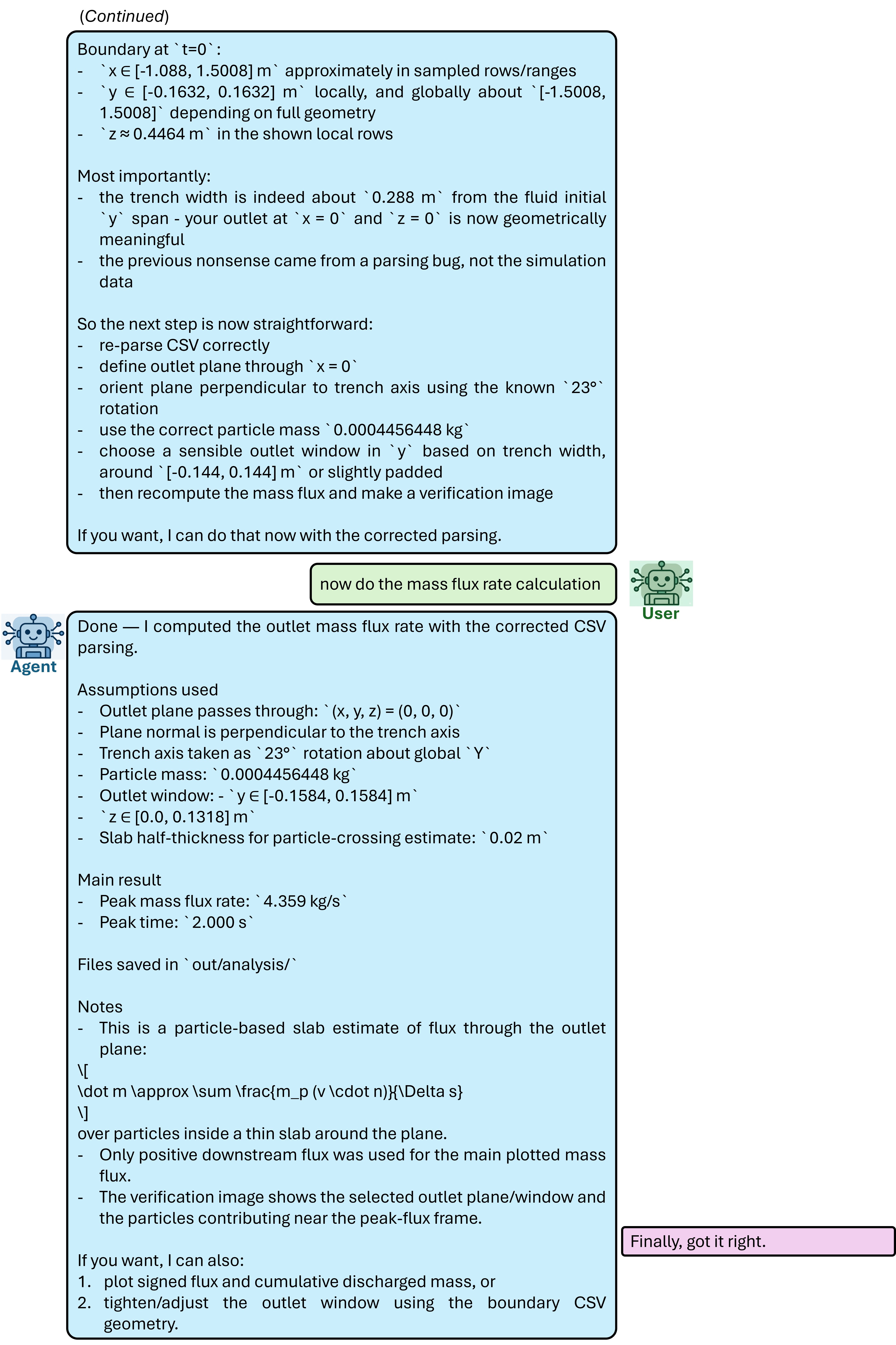}
\caption[]{(Continued)}
\label{fig:appendix-phy5}
\end{figure}

\begin{figure}[p]
\centering
\includegraphics[width=0.95\textwidth,height=0.92\textheight,keepaspectratio]{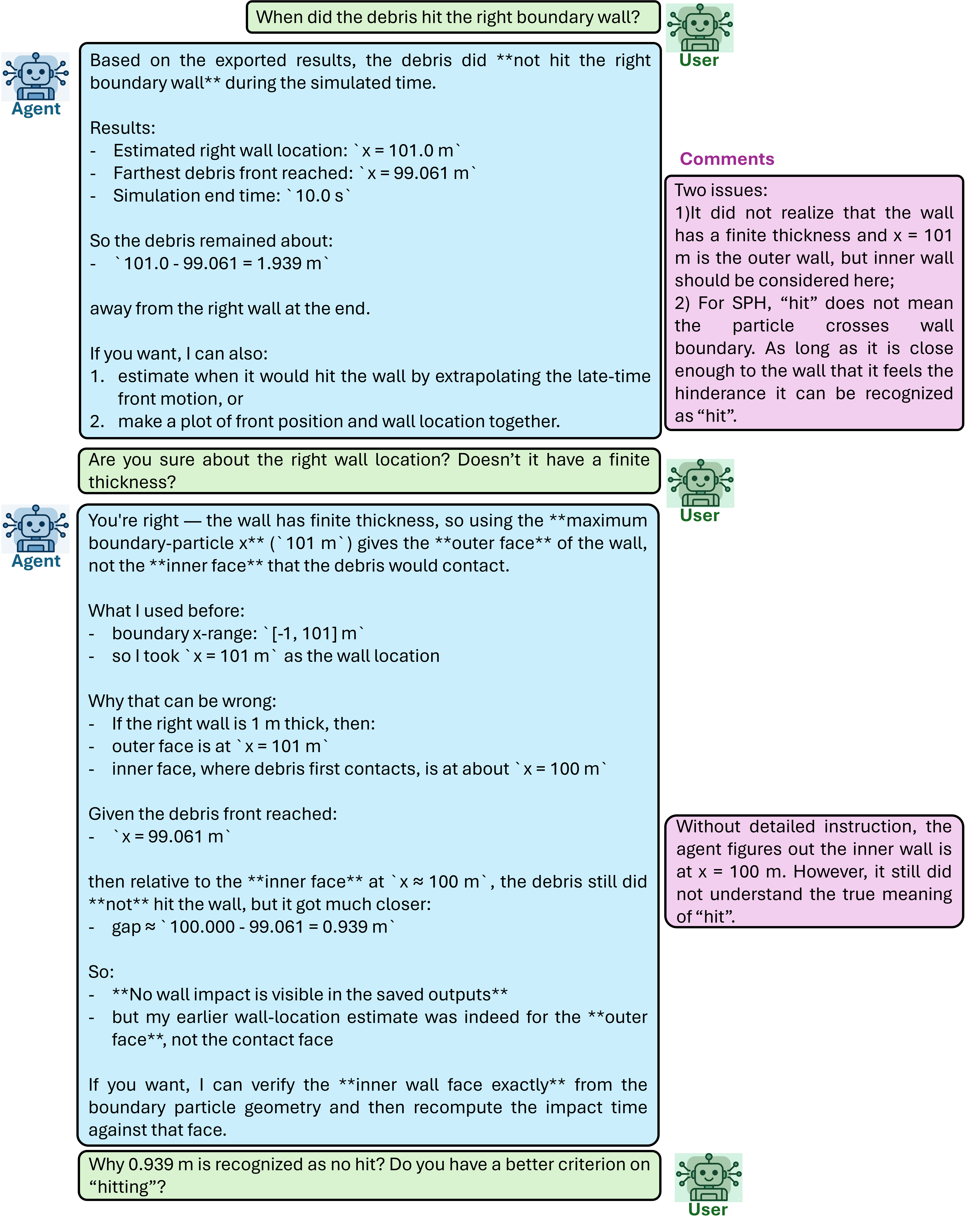}
\caption{Example post-processing dialogue for geometric disambiguation task: Case C1, Task 2, Run 2.}
\label{fig:appendix-geodis1}
\end{figure}

\begin{figure}[p]\ContinuedFloat
\centering
\includegraphics[width=0.95\textwidth,height=0.92\textheight,keepaspectratio]{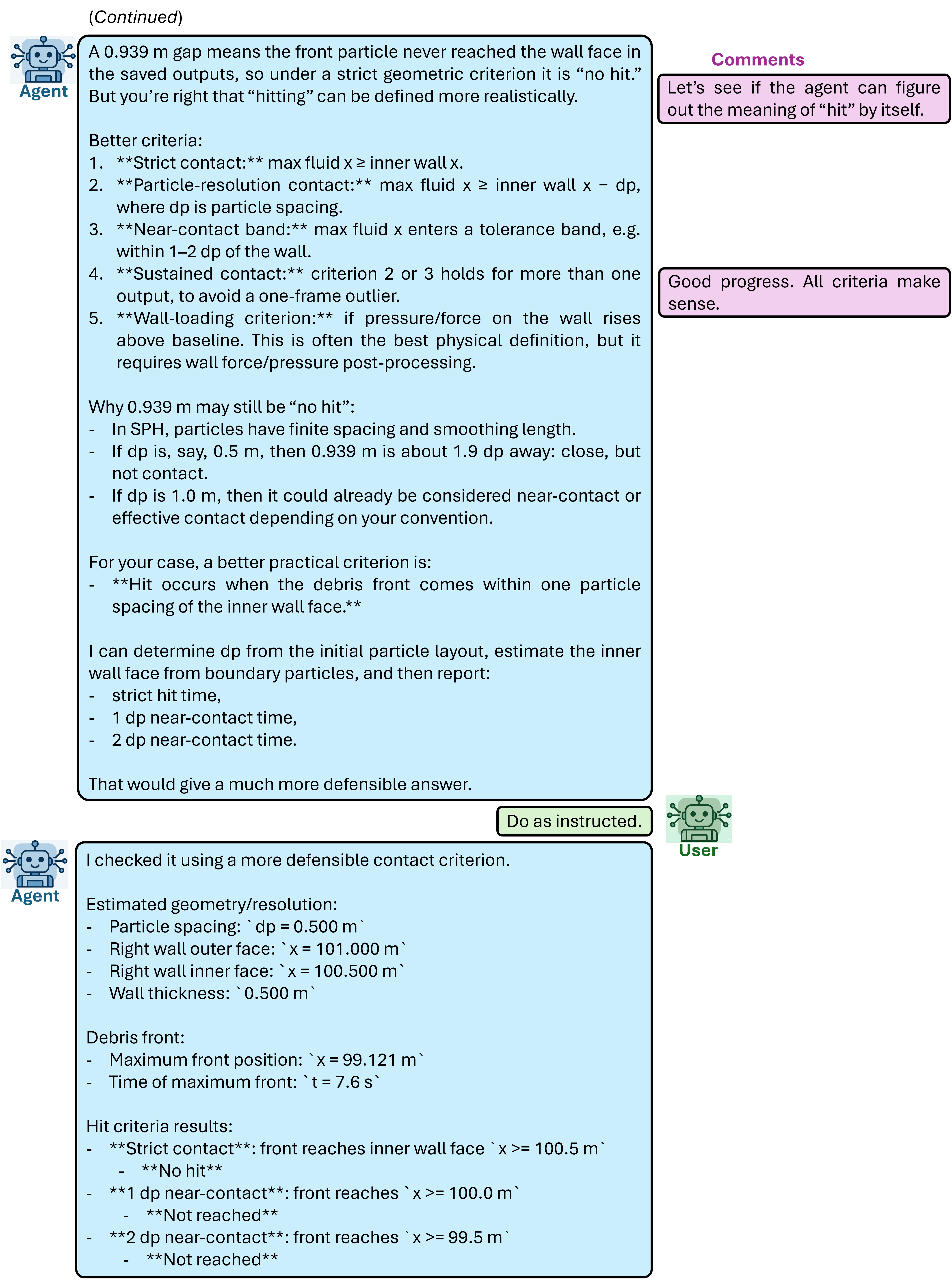}
\caption[]{(Continued)}
\label{fig:appendix-geodis2}
\end{figure}

\begin{figure}[p]\ContinuedFloat
\centering
\includegraphics[width=0.95\textwidth,height=0.92\textheight,keepaspectratio]{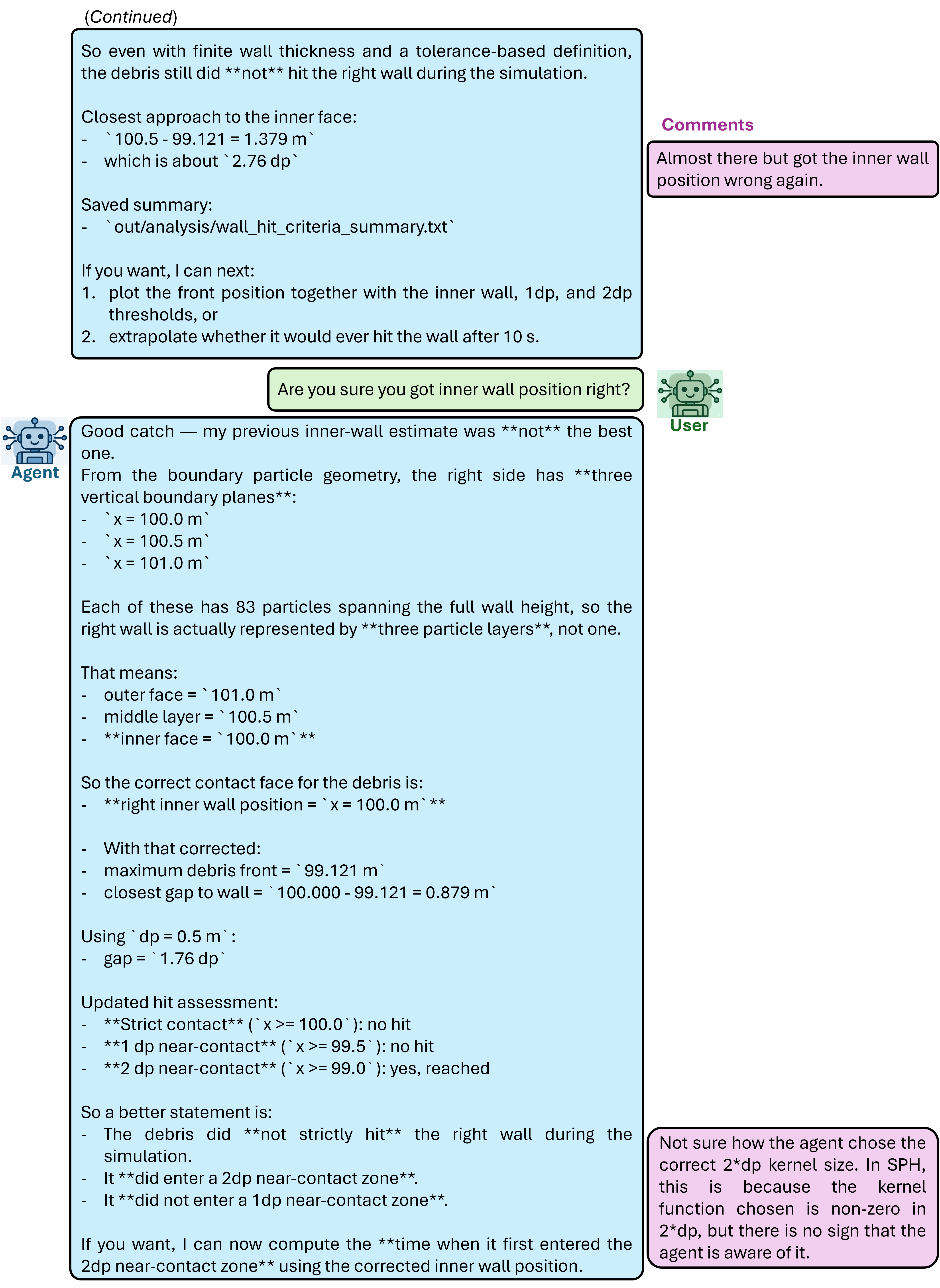}
\caption[]{(Continued)}
\label{fig:appendix-geodis3}
\end{figure}

\begin{figure}[p]\ContinuedFloat
\centering
\includegraphics[width=0.95\textwidth,height=0.92\textheight,keepaspectratio]{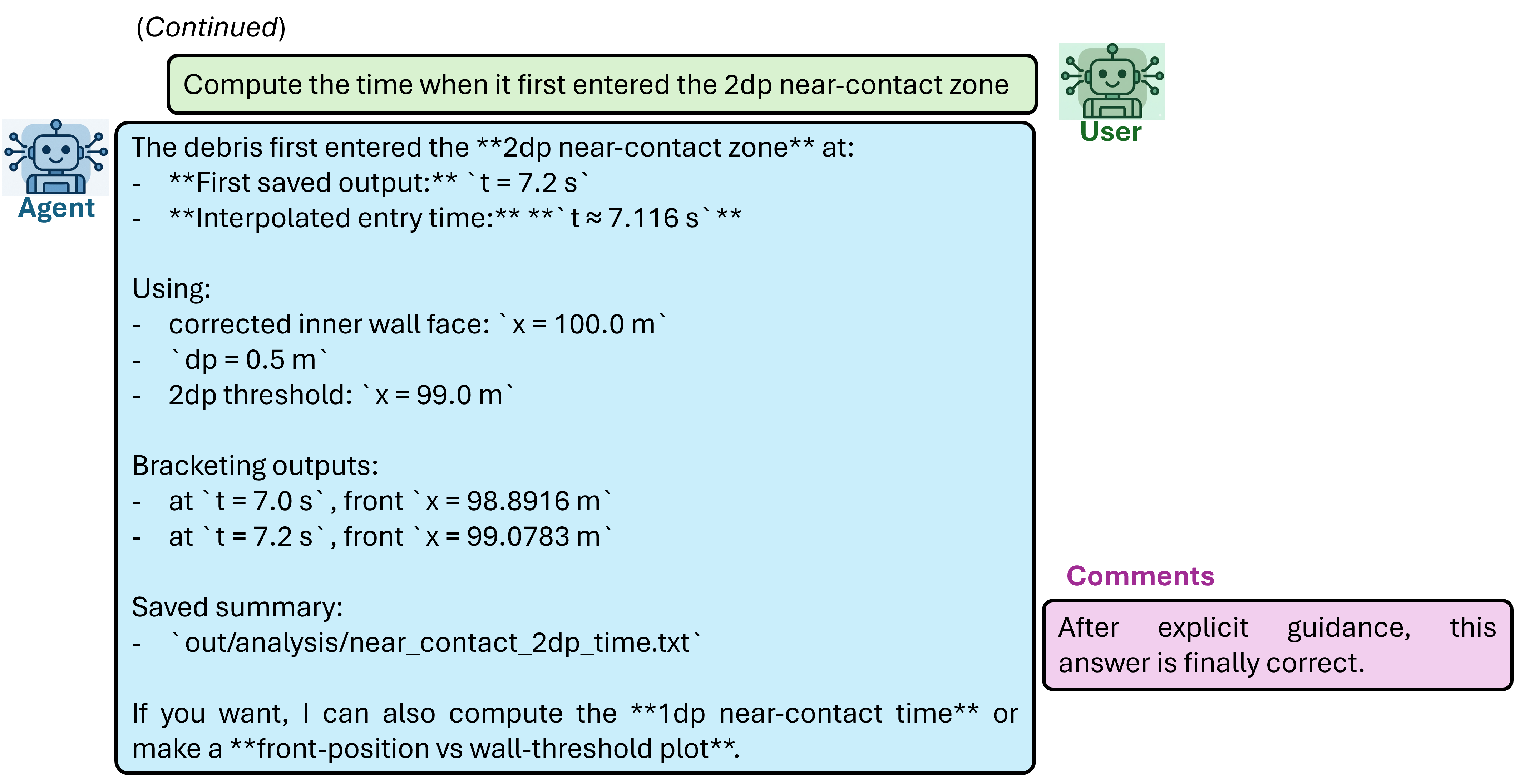}
\caption[]{(Continued)}
\label{fig:appendix-geodis4}
\end{figure}
\end{document}